\documentclass[twocolumn,amsmath,showkeys,amssymb,superscriptaddress,longbibliography,floatfix, pre]{revtex4-2}

\pdfoutput=1

\usepackage{bm}
\usepackage{yfonts}
\usepackage{graphicx,epsfig}% Include figure files
\usepackage{amsmath,amssymb,mathtools}%,amsfonts,multirow,rotate,color}amssymb
\usepackage[rgb]{xcolor}
\usepackage{hyperref}
\usepackage{braket}
\def\be{\begin{equation}}
\def\ee{\end{equation}}

\def\bc{\begin{center}}
\def\ec{\end{center}}
\def\bea{\begin{eqnarray}}
\def\eea{\end{eqnarray}}

\newcommand{\Avg}[1]{\left\langle{#1}\right\rangle}

\begin{document}

\title{Gravity from entropy}

\author{Ginestra Bianconi}
\email{ginestra.bianconi@gmail.com}
\affiliation{School of Mathematical Sciences, Queen Mary University of London, London, E1 4NS, United Kingdom}

\begin{abstract}
Gravity is derived from an entropic action coupling matter fields with geometry. The fundamental idea is to relate the metric of Lorentzian spacetime  {to a quantum operator, playing the role of an {\em renormalizable effective density matrix} }and to describe the matter fields topologically, according to  a Dirac-K\"ahler formalism, as the direct sum of a  zero-form, a one-form and a two-form. While the geometry of spacetime is defined by its metric, the matter fields can be used to define an alternative metric, the metric induced by the matter fields, which geometrically describes the interplay between spacetime and  matter.  The proposed entropic action is the quantum relative entropy between the metric of spacetime and the metric induced by the matter fields. The modified Einstein equations  obtained from this action reduce to the Einstein equations with zero cosmological constant in the regime of low coupling. By introducing  the {\em G-field}, which acts as a set of Lagrangian multipliers, the proposed entropic action reduces to a dressed Einstein-Hilbert action with an emergent small and positive cosmological constant only dependent on the G-field. The obtained equations of modified gravity remain second order in the metric and in the G-field. A canonical quantization of this field theory could bring new insights into quantum gravity while further research might clarify the role that the G-field could have for dark matter.
\end{abstract}
\maketitle
\section{Introduction}
The relation between general relativity, statistical mechanics and information theory is a central research topic in theoretical physics. The interest in the subject has its roots in the discovery that black holes have an entropy \cite{bekenstein1973black,bekenstein1974generalized} and emit Hawking radiation~\cite{hawking1975particle}. 
Recently, important results have been obtained  relating information theory, entanglement entropy \cite{ryu2006aspects,nishioka2009holographic,faulkner2013quantum}  and gravity \cite{jacobson1995thermodynamics,carroll2016entropy,jacobson2016entanglement,callan1994geometric,chirco2014spacetime,chirco2010nonequilibrium,verlinde2011origin}
involving the holographic principle \cite{hooft2001holographic,susskind1995world,swingle2012entanglement}, the   entanglement properties of quantum field theory and the theory of von Neumann algebras~\cite{witten2018aps,sorce2023notes}. 

These results define a very active research direction \cite{padmanabhan2010thermodynamical} indicating that the quest for an ultimate gravitational theory based on information theory and statistical mechanics is ongoing. A comprehensive statistical mechanics approach to gravity  is expected to give rise to modified Einstein equations~\cite{sotiriou2010f,sotiriou2007metric,woodard2007avoiding} that on the one side can be testable experimentally \cite{berti2015testing,barausse2020prospects} while on the other side  can bring important conceptual insights into the ultimate theory for black holes \cite{barack2019black}, dark matter \cite{bertone2005particle} and   quantum gravity \cite{Rovelli,CDT,Rahmede,eichhorn2019asymptotically,Oriti,barcelo2011analogue}.

In this work,  a continuum modified theory of gravity based on a statistical mechanics action is considered. This theory treats the metric at each point of spacetime as a ``renomalizable" density matrix, or more precisely a local quantum operator.  {  This central idea is relating geometry with  the mathematical foundation of quantum field theory~\cite{araki1999mathematical} and is inspired by the used of von Neumann algebras in explaining  entanglement in field theories \cite{witten2018aps,sorce2023notes} and quantum gravity ~\cite{ciolli2020information,longo2021neumann,witten2022gravity}.
}
While the geometry of spacetime is defined through its associated metric, the interplay between matter fields and geometry is captured by the metric induced by the matter fields which describes how the matter fields effectively curve spacetime.
Embracing a statistical mechanics approach to gravity,  this work   { interprets these metric tensors as quantum operators and  postulates  an action for gravity given by  the quantum relative entropy between the metric of the manifold and the metric  induced by the matter fields.}    {From the mathematical point of view, the  quantum relative entropy proposed in this work is strictly related  to the Araki quantum relative entropy for von Neumann algebras~\cite{araki1975relative,
ohya2004quantum,witten2018aps,vedral2002role}.
From the physics point of view,  the proposed action  fully describes how matter curves geometry and how geometry affects the matter fields.}

A crucial aspect of the proposed  theory is the adoption of a topological (Dirac-K\"ahler like \cite{kruglov2002dirac,becher1982dirac}) description of bosonic matter fields.
Note that  the extension of  Dirac-K\"ahler and staggered fermions formalism to  bosonic particles  is gaining increasing interest in lattice gauge theory~\cite{berenstein2024staggered,berenstein2023staggered} and in network theory as well~\cite{bianconi2024quantum}.  
These bosonic matter fields are described as the direct sum of a $0$-form, a $1$-form, and a $2$-form  defined on the Riemannian manifold describing spacetime. Moreover the metric induced by the topological matter fields  is expressed in terms of the Hodge-Dirac operator \cite{nakahara2018geometry}.
From this statistical mechanics approach to gravity we derive the modified Einstein equations by introducing an auxiliary  field associated to gravity which we call {\em G-field}. The introduction of this new field is justified as it acts as a set of Lagragian multipliers enforcing linear constraints on metric induced by the matter fields. In this way the G-field  extends the popular use of the Legendre transformation in $f(R)$ theories~\cite{woodard2007avoiding,sotiriou2010f}.
Given the particular entropic structure of the action, the modified Einstein equations take a very simplified expression. The gravitational part of the action takes the form of a {\em dressed Einstein-Hilbert} action in which we observe an emergent positive cosmological constant that depends exclusively on the G-field.
%The modified equations generalize results ~\cite{woodard2007avoiding,sotiriou2010f} on the $f(R)$ theory and avoid the Ostrogradsky instability as they contain only second derivatives of the metric and of the G-field.

This work greatly expands on previous results~\cite{bianconi2024quantum} obtained in the discrete setting by the same author. On one side, here a continuum and fully Lorentz invariant theory is proposed. This progress is based on the development of the suitable mathematical framework to define the Lorentz invariant entropy and cross-entropy between the metric of the spacetime and the metric induced by the matter fields.
On the other side, here the relation of this statistical mechanics/information theory action with the Einstein-Hilbert action \cite{einstein1915feldgleichungen,carroll2019spacetime} is established defining a clear connection to gravity. Two fundamental aspects of this work that are not present in Ref. \cite{bianconi2024quantum}, are relevant to clearly relate this approach to gravity. First, this work  considers a local theory, defining the entropy of the metric at each point of spacetime, while the previous work considers only the entropy associated to the full metric of the higher-order network. The present local theory allows a closer connection to gravity and  constitutes a step forward to establish the connection between this approach and the quantum theory of entanglement~\cite{witten2018aps}. Secondly, by adopting the continuum limit, in this work  the intrinsic difficulty related to the definition of the curvature of networks, simplicial and cell complexes is avoided.

To keep the discussion concise the focus is here mostly on scalar (bosonic) matter fields,  and their topological generalizations  {with a brief mention in Appendix $\ref{Ap_gauge}$ to the natural  extension of this framework to Abelian gauge fields,  while in Ref.~\cite{bianconi2024quantum} the theory covers also fermionic matter fields}. Further extensions of the proposed  local framework to Dirac and non-Abelian gauge fields \cite{bianconi2024quantum,bianconi2023mass,delporte2023dirac}  in the continuum or in the discrete setting are left for future investigations.\\

This work is structured as follows. In Sec. \ref{sec:preliminaries}  we  provide the motivation of the proposed theory  and we discuss  preliminary results on an instructive warm-up scenario.
In Sec. \ref{sec:full} we outline the proposed theoretical framework, we postulate the entropic action for gravity and we derive the corresponding modified Einstein equations. In Sec.\ref{sec:conclusions} we provide the concluding remarks. The paper also includes three Appendices discussing possible extensions of the proposed theoretical framework, providing the mathematical background and all the necessary details regarding the notation used Sec. \ref{sec:full}, {  and establishing the connection of the present theory with the theory of local quantum operators and the Araki entropy.} 
\section{Motivation of the theory and preliminary considerations}
\label{sec:preliminaries}
\subsection{Eigenvalues and logarithm  of rank 2-tensors}
\label{sec:mt}
Spacetime is described by a torsion free, $d$-dimensional Riemannian manifold $\mathcal{K}$ associated with a Lorentzian metric $g_{\mu\nu}$ of signature $\{-1,1,,1\ldots, 1\}$ and a metric compatible  Levi-Civita connection
$\Gamma_{\nu\mu}^{\sigma}$ determining the covariant derivative $\bm\nabla_{\mu}$.
In order to formulate our statistical mechanics and information theory action for gravity, we need first to define the eigenvalues the  logarithm of  rank-$2$ tensors $\hat{\bf G}$. 
To this end we first define the eigenvalues $\lambda$ and eigenvectors  $V^{(\lambda)}_{\nu}$ of the covariant tensor $\hat{\bf G}$ of elements $\hat{\bf G}_{\mu\nu}$ in a Lorentz invariant way. These  satisfy the eigenvalue  problem
\bea
\hat{G}_{\mu\nu}[V^{(\lambda)}]^{\nu}=\lambda V^{(\lambda)}_{\mu}.
\label{eig1}
\eea
We say that a  rank-2 tensor is positively defined if all its eigenvalues are positive.
We notice that this definition of the eigenvalue of a rank-2 tensor reduces to the definition of the eigenvalue of the matrix $\hat{\bf G}{g}^{-1}$ as Eq.(\ref{eig1}) can be rewritten as
\bea
\hat{G}_{\mu\sigma}g^{\sigma\nu}[V^{(\lambda)}]_{\nu}=\lambda V^{(\lambda)}_{\mu}.\label{eig2}
\eea
One striking consequence of this definition is that   the eigenvalues of the metric $g_{\mu\nu}$ are all identically equal to one.

Assuming that the  tensor $\hat{G}_{\mu\nu}$ is positively defined,  we  define the logarithm of this tensor as 
\bea
[\ln(\hat{\bf G})]_{\mu\nu}=V_{\mu}^{(\lambda)}V^{(\lambda)}_{\nu}\ln(\lambda).
\eea
and the inverse of a tensor as
\bea
[\hat{\bf G}^{-1}]^{\mu\nu}=[V^{(\lambda)}]^{\mu}[V^{(\lambda)}]^{\nu}\lambda^{-1}.
\eea
It  follows that if the tensor $\hat{\bf G}$ is invertible, the logarithm of the inverse of a positively defined tensor   is given by 
\bea
[\ln(\hat{\bf G}^{-1})]^{\mu\nu}=-[V^{(\lambda)}]^{\mu}[V^{(\lambda)}]^{\nu}\ln(\lambda).
\eea
Finally we define the trace of a rank two tensor as the sum of its eigenvalues, i.e.
\bea
\mbox{Tr} \hat{\bf G}=\sum_{\lambda}\lambda.
\eea
Thus the trace of a rank-$2$ tensor can be also calculated  as usual in tensor calculus, as the trace of the matrix $\hat{\bf G}g^{-1}$, i.e.
\bea
\mbox{Tr} \hat{\bf G}=\mbox{Tr}_M\hat{\bf G}g^{-1}=\hat{G}_{\mu\nu}g^{\mu\nu}.
\eea
In this first section we are interested exclusively on rank-two tensors that are metrics between  {vectors (and 1-forms)}. In the subsequent paragraphs and in Appendix \ref{ApA} we will extended the notion of eigenvalues also to metric matrices between  {bivectors (and 2-forms)} represented by rank-4 tensors. Such an approach will be shown in Appendix \ref{ApA} to be general and  applicable to metric tensors between  two   {$n$-vectors (an $n$-forms)} with $n$ of any order. 

\begin{table*}
\begin{tabular}{c cc}
\hline
\hline
Covariant Metrics  \qquad & Elements \qquad &  Definition\\
\hline
 $g$		&$g_{\mu\nu}$			& Default covariant rank-2 metric tensor between vectors associated to the manifold $\mathcal{K}$ \\
  ${\bf G}$		&$G_{\mu\nu}$			& Covariant rank-2 metric between vectors induced by the matter fields\\
    $\hat{\bf G}$	&	$\hat{G}_{\mu\nu}$			& General covariant rank-2 metric tensor  including both $g$ and ${\bf G}$\\
 \hline
 \hline
\end{tabular}
\caption{ {Notation used to indicate the different metrics in the warm-up-scenario}}
\label{table_warm_up_scenario}
\end{table*}

\subsection{Warm-up scenario: Entropic action}
Before we develop our theory, let us consider an instructive warm-up scenario that will help justify the theory that we will present in the following. 
Having defined the logarithm of  positively defined rank-2 tensors, we are now in the position to define their Lorentz invariant quantum entropy $H$ for metric associated to $1$-forms. This is inspired by the expression of the Von Neumann entropy albeit we do not require the tensor to have trace one at every point in spacetime.  {Thus strictly speaking we interpret the  generic metric tensor as a quantum operator~\cite{araki1999mathematical}, which has the physical interpretation of a {\em renormalizable effective density matrix}~\cite{sorce2023notes}.}
We consider  positive definite, invertible metric for $1$-forms, represented by the covariant tensor $\hat{\bf G}$ of rank two.
We define the entropy $H$ of $\hat{\bf G}$  as 
\bea
H=\mbox{Tr}\hat{\bf G}\ln\hat{\bf G}^{-1}=\hat{G}_{\mu\nu}[\ln \hat{\bf G}^{-1}]^{\nu\mu}=-\sum\lambda\ln \lambda.
\eea
Note that the problem of defining eigenvalues and entropy of tensors is a topic of intense research and similar definitions have been provided for instance in the theory of elasticity \cite{de2011computational} and in applied tensor analysis \cite{qi2017tensor} (although  not defined in a Lorentz invariant way).
 {In the present  warm-up scenario we will define the entropy and the quantum relation entropy of different metrics taking the form of rank-2 tensors. As a reference for the notation used, the reader can refer to Table \ref{table_warm_up_scenario}.}
Recalling that the metric $g$  has all the eigenvalues equal to one, it follows that the entropy of  the metric  $g$ is null, i.e. 
\bea
H=\mbox{Tr}{g}\ln{g}^{-1}=0.
\label{H}
\eea
The fundamental  assumption  of the present theory is that spacetime is endowed with two metrics: the metric $g$ fully determining the spacetime geometry and the metric induced by the matter fields ${\bf G}$ that fully capture the interplay between matter fields and geometry. Leaving the detailed discussion about the metric induced by the matter fields to the next paragraph, we postulate  that the action should explicitly express the relation between these two metrics and their reciprocal coupling.
By embracing a statistical mechanics approach we  consider the Lagrangian $\mathcal{L}$  given by the
 quantum relative entropy between the metric $g$ and the metric ${\bf G}$ induced by the matter field  defined as 
\bea
\mathcal{L}&=&-\mbox{Tr}{g}\ln{g}^{-1}+\mbox{Tr}{g}\ln{\bf G}^{-1}.
\label{action}
\eea
Using $H=0$ we  thus obtain:
\bea
{\mathcal{L}}=\mbox{Tr}g\ln {{\bf G}}^{-1}.
\label{L}
\eea
We observe that since $g$ has all the eigenvalues equal to the identity,   and Eq.(\ref{eig2}) holds for $\hat{\bf G}={\bf G}$,   the Lagrangian $\mathcal{L}$ can be also be expressed as 
\bea
\mathcal{L}\equiv-\mbox{Tr}_M\ln \ {\bf G}{g}^{-1}=-\sum_{\lambda^{\prime}}\ln(\lambda^{\prime}),
\label{L2a}
\eea
where with  $\mbox{Tr}_M \ln \ {\bf G}{g}^{-1}$ we indicate the usual trace of the logarithm of the matrix  $\ln \ {\bf G}{g}^{-1}$ and with $\lambda^{\prime}$ the eigenvalues of the $2$-tensor ${\bf G}$ as defined in Eq.(\ref{eig2}).  

An  important point here is that the Lorentz invariant definition of the quantum relative entropy given by Eq.(\ref{L2a})  requires that the matrix ${\bf G}$ has  to be invertible as well as the matrix $g$. 
Thus this is another difference with respect to the strictly speaking density matrices that can be semi-definite positive.
We are now in the position to  consider an action $\mathcal{S}$ associated to the Lagrangian $\mathcal{L}$, as given by  
\bea
\mathcal{S}=\frac{1}{\ell_P^d}\int \sqrt{|-g|}\mathcal{L} d{\bf r},
\eea
where $|-g|$ indicates the absolute value of the determinant of $g$ and $\ell_P$ indicates the Planck length. \\

 {We note that the quantum relative entropy \cite{araki1975relative,ohya2004quantum,witten2018aps} is a central quantity in quantum information~\cite{vedral2002role}, in  the theory of local quantum  operators~\cite{araki1999mathematical}
and the mathematical foundations of quantum gravity ~\cite{ciolli2020information,longo2021neumann,witten2022gravity}.}

 {Although we are not aware of previous interpretations of the metrics as quantum operators,  it is well known that the quantum relative entropy can also be defined among quantum operators that generalize density matrices. 
 %In particular the Araki quantum relative entropy~\cite{araki1975relative,ohya2004quantum}.
% is defined between quantum operators, that like our metric matrices admit a  finite trace at each point of the manifold but this trace might  be not unitary and be a function of the considered point of the manifold. 
{ In particular, in the fundamental theory of quantum operators algebras~\cite{araki1975relative,ohya2004quantum}, quantum operators generalize the notion of density matrices in an analogous way of the metric tensors adopted here. In particular in this theory, quantum operators, like our metric matrices, might admit a  finite trace at each point of the manifold but this trace might  be not unitary and might be a function of the considered point of the manifold. For these quantum operators  the Araki  quantum relative entropy~\cite{araki1975relative,ohya2004quantum}
  is defined. This entropy reduces to the von Neumann entropy when the quantum operators reduces to a density matrices of unitary trace but is defined also among quantum operators of non unitary trace. }
 }

 {While we leave the discussion of the relation of our entropic action to the Araki quantum relative entropy to Appendix \ref{quantum}, a series of  physical observations are here in place to justify our interpretation of the metrics matrices as quantum operators. 
As remarked previously we treat the metric tensors as quantum operators or effective density matrices. The main differences between   the metric tensor and the density operators  include the fact that we require that the metric matrices are invertible and we do not require that they have unitary trace at each point of the manifold.}
The requirement of treating exclusively invertible metrics $g$ and ${\bf G}$ is dictated by our desire to define the entropy in a Lorentz invariant way.  In  fact here we desire to treat the metrics $g$ and ${\bf G}$ on the same footing of their inverse $g^{-1}$ and ${\bf G}^{-1}$.
It might be argued that also our relaxation of the requirement to have density matrices of trace one, has similar roots.  In fact if we require $g$ and  ${\bf G}$  to have unitary trace at each point of the manifold their  inverse as well as their dual (defined in Appendix $\ref{quantum}$) in general  will not have a unitary trace.

\subsection{Warm-up scenario: scalar matter fields}
We now apply this warm-up scenario in presence of scalar matter fields.
We consider the  complex valued scalar matter field $\phi({\bf x})\in \mathbb{C}$ with   {${\bf x}\in {\mathcal{K}}$} and we consider  the $d$-dimensional manifold immersed in   $\mathcal{K}\otimes \mathbb{C}$ defined by the points $({\bf x},\phi({\bf x}))$. 
The metric
${\bf G}$ induced on $\mathcal{K}$ by this matter fields is given by  
  \bea
{\bf G}=g+\alpha {\bf M},
\label{metricinduced}
\eea
where  $\alpha$ is a real positive  parameter,  ${\bf M}$ is the rank-2 tensor of elements 
\bea
M_{\mu\nu}=({\bm\nabla}^{\mu}\bar\phi)({\bm\nabla}^{\nu}\phi),
\eea 
where here and in the following $\bar{\dot}$ indicates complex conjugation.  Since ${\bf G}$ should be adimensional it is convenient to work in the units $\hbar=c=1$ and to  put $\alpha=\alpha'\ell_P^d$ where $\ell_P$ is the Planck length and $\alpha'$ is adimensional. 
 
We observe that in the limit in which the field $\phi$ is real, i.e. $\phi({\bf x})\in \mathbb{R}$ and the metric flat and Euclidean, i.e. $g_{\mu\nu}=\delta_{\mu\nu}$, the metric induced by the matter fields ${\bf{G}}$ given by Eq.(\ref{metricinduced}) reduces to the first fundamental form of Gauss for the $d$ manifold immersed in $\mathbb{R}^{d+1}$ defined by the set of points $({\bf x},\phi({\bf x}))$.
For the underlying mathematical treatment of the metric induced by  real scalar fields  we refer the interested reader to  Ref. \cite{keski2022mathematical}. 
 
Let us define the scalar product $|{\bm\nabla}\phi|^2$ as 
\bea
|{\bm\nabla}\phi|^2={\bm\nabla}_{\mu}\bar\phi{g}^{\mu\nu}{\bm\nabla}_{\nu}\phi.
\eea

We observe that the inverse ${\bf G}^{-1} $ of the induced metric ${\bf G}$ has metric  given by
\bea
[{{\bf G}^{-1}}]^{\mu\nu}=g^{\mu\nu}-\alpha\frac{M^{\mu\nu}}{1+\alpha|{\bm \nabla}{\phi}|^2}.
\eea
Adopting this notation we  observe that  the logarithm of the induced metric $\ln {\bf G}$ and the logarithm of $\ln {\bf G}^{-1}$ have elements 
\bea 
\left[\ln {\bf G}\right]_{\mu\nu}&=&f(|{\bm \nabla}\phi|^2){M}_{\mu\nu},\nonumber \\
\left[\ln {\bf G}^{-1}\right]^{\mu\nu}&=&-f(|{\bm \nabla}\phi|^2){M}^{\mu\nu},
\label{lnG}
\eea
where
\bea
f(w)=\frac{\ln(1+\alpha w)}{w}.
\eea
where $f(w)\to \alpha$ for $ |\alpha w|\ll1$. 
With this choice of the metric induced by the matter field we obtain that  the Lagrangian $(\ref{L})$ reads
\bea
{\mathcal{L}}&=&-\ln(1+\alpha{|{\bm \nabla}{\phi}|^2}).
\label{L1b}
\eea
We observe that in the limit $\alpha|\nabla \phi|\to 0$ then we have 
$\mathcal{L}\to -\alpha|{\bm \nabla}{\phi}|^2$, i.e. we recover the Lagrangian corresponding to the massless Klein-Gordon equation.

By minimizing the action $\mathcal{S}$ with respect to the field $\phi$ and to the metric $g$ we obtain the Euler-Lagrange equations of motion:
\bea
\bm\nabla_{\mu}h(|\nabla\phi|^2)g^{\mu\nu}\bm\nabla_{\nu} \phi&=&0,
\label{dif2}
\eea
where $h(w)$ is given by
\bea
h(w)=\frac{\alpha}{1+\alpha w}.
\eea
In the limit $\alpha w\to 0$ we have $h(w)\to \alpha$ and the equation for the scalar field reduces to the simple  massless Klein-Gordon equation.
%As we have already discussed previously, one might wonder whether our choice to take directly the metric matrix and interpret it as a ``renormalizable density matrix" is the most proper one, and whether it would be better to   instead renormalize directly $g$ and ${\bf G}$ with their trace.
%As an additional  possible answer to this important question, in Sec. \ref{quantum} we  will relate our entropic action to the Araki quantum relative entropy of von Neumann algebras.
By putting equal to zero the variation of the action $\delta \mathcal{S}=0$ with respect to the metric $g$,  we get 
\bea
\delta\mathcal{S}=-h(|\nabla\phi|^2)M^{\mu\nu}-\frac{1}{2}\mathcal{L}g^{\mu\nu}=0.
\eea
In empty spacetime ${\bf M}=0$ and $\mathcal{L}=0$, thus this equation is automatically satisfied independently on the value of  $g$. Thus in this limit, the metric $g$ is not determined by the action.\\
Let us now make some remarks about this warm-up derivation.
We proposed a statistical mechanics   framework that is very inspiring as we get the massless Klein-Gordon equation as the outcome of the minimization of  a quantum entropy action for low coupling, i.e. $0<\alpha|\nabla \phi|^2\ll 1$. However this approach has two important limitations. The first limitation is that  the Klein-Gordon equation does not contain the mass term.  The second limitation is that in  absence of matter fields the metric is not determined.
In the following we provide a more comprehensive  framework to address these two limitations. This framework can be  related  to gravity as it gives rise to modified Einstein equations that reduce to the Einstein equations for low coupling.

\section{The entropic theory of matter fields coupled to geometry}
\label{sec:full}

\begin{table*}
\begin{tabular}{c c}
\hline
\hline
Covariant Topological Metrics  \qquad &  Interpretation\\
\hline
 $\tilde{g}=1 \oplus g_{\mu\nu} dx^{\nu}\otimes dx^{\nu}\oplus {[g_{(2)}]}_{\mu\nu\rho\sigma}(dx^{\mu}\wedge dx^{\nu})\otimes(dx^{\rho}\wedge dx^{\sigma})$					& Default covariant topological metric between \\
 & the topological fields associated to the manifold $\mathcal{K}$\\
  $\tilde{\bf G}=G_{(0)}\oplus[G_{(1)}]_{\mu\nu}dx^{\mu}\otimes dx^{\nu}\oplus [G_{(2)}]_{\mu\nu\rho\sigma}(dx^{\mu}\wedge dx^{\nu})\otimes(dx^{\rho}\wedge dx^{\sigma})$				&Covariant topological metric induced by the matter fields\\
  \hline\hline
  \end{tabular}
  \caption{ {Covariant topological metrics used in the general scenario.}}
\label{table_general_scenario_I}
\end{table*}
\begin{table*}
  \begin{tabular}{ccc}
 \hline
 \hline
 Covariant metric tensors  & Elements &  Intepretation\\
 \hline
 $1$&$1$& Default metric tensor between scalars \\
 $g$ & $g_{\mu\nu}$ & Default metric tensor between vectors \\
 $g_{(2)}$& ${[g_{(2)}]}_{\mu\nu\rho\sigma}$ &Default metric tensor between bivectors\\
  ${ G}_{(0)}$ & ${ G}_{(0)}$ &  Metric tensor between scalars induced by the matter fields\\
 ${\bf G}_{(1)}$ &${[{\bf G}_{(1)}]}_{\mu\nu}$ &Metric tensor between vectors induced by the matter fields\\
 ${\bf G}_{(2)}$ & ${[{\bf G}_{(2)}]}_{\mu\nu\rho\sigma}$ &Metric tensor between bivectors induced by the matter fields\\
 \hline
 \hline
\end{tabular}
\caption{ {Covariant metric tensors between $n$-vectors that are included in the covariant topological metrics.}}
\label{table_general_scenario_II}
\end{table*}

\subsection{Topological matter fields and their associated metrics}

In order to derive gravity from our entropic action we need to consider the topological bosonic matter field.
The topological bosonic matter field is a type of Dirac-K\"ahler \cite{kruglov2002dirac,becher1982dirac}  boson given by the  direct sum of a 0-form, a 1-form and a 2-form. Topological bosonic fields are receiving increasing attention in discrete theories developed in network \cite{bianconi2024quantum} and lattice gauge theories \cite{berenstein2024staggered,berenstein2023staggered}.
Taking into consideration topological bosonic fields will allow us to introduce in the metric induced by the matter fields, terms depending on the mass of the bosonic field. Thus, in this way, we address the first limitation of the warm-up scenario that we have presented above.

In order to address the second limitation of the warm-up scenario discussed previously,  we  include in the expression of the metric induced by the matter field, also terms depending directly on the curvature of the manifold. These terms will be expressed in terms of the Ricci scalar $R$, the Ricci (covariant) tensor ${\bf R}$ of elements $R_{\mu\nu}$ and the Riemann tensor of elements $R_{\mu\nu\rho\sigma}$. 
 
We define the topological fields $\ket{\Phi}$ as the direct sum between a zero (complex valued) form $\phi$ and a (complex valued) one-form $\omega_{\mu}dx^{\mu}$, and a (complex valued) two-form $\zeta_{\mu\nu}dx^{\mu}\wedge dx^{\nu}$, with $\zeta_{\mu\nu}=-\zeta_{\nu\mu}$ i.e.
\bea
\ket{\Phi}=\phi\oplus\omega_{\mu} dx^{\mu}\oplus\zeta_{\mu\nu}dx^{\mu}\wedge dx^{\nu},
\eea
and its conjugate topological field $\bra{\Phi}$ as
\bea
\bra{\Phi}=\bar{\phi}\oplus\bar{\omega}_{\mu} dx^{\mu}\oplus\bar{\zeta}_{\mu\nu}dx^{\mu}\wedge dx^{\nu}.
\eea
The considered covariant metric  $\tilde{g}$ is defined as the direct sum of the metric   among  {scalars} (the identity), the metric  $g$  {among vectors} introduced previously, and the  metric $g_{(2)}$  {among bivectors} given by 
\bea
g_{(2)}=g_{\mu\rho}g_{\nu\sigma}(dx^{\mu}\wedge dx^{\nu})\otimes(dx^{\rho}\wedge dx^{\sigma}),\eea
or, exploiting the anti-symmetry of the $2$-forms,
\bea
g_{(2)}={[g_{(2)}]}_{\mu\nu\rho\sigma}(dx^{\mu}\wedge dx^{\nu})\otimes(dx^{\rho}\wedge dx^{\sigma}),
\eea
with 
\bea
{[g_{(2)}]}_{\mu\nu\rho\sigma}=\frac{1}{2}(g_{\mu\rho}g_{\nu\sigma}-g_{\mu\sigma}g_{\nu\rho}).
\label{g2}
\eea
It then follows that $\tilde{g}$ is given by  
\bea
\tilde{g}&=&1 \oplus g_{\mu\nu} dx^{\nu}\otimes dx^{\nu} \\
&&\oplus {[g_{(2)}]}_{\mu\nu\rho\sigma}(dx^{\mu}\wedge dx^{\nu})\otimes(dx^{\rho}\wedge dx^{\sigma}).\nonumber
\eea
The  {local} scalar product among topological fields is  defined as
\bea
\Avg{\Phi|\Phi}= |\phi|^2+\bar{\omega}^{\mu}\omega_{\mu}+\bar{\zeta}^{\mu\nu}\zeta_{\mu\nu},
\eea
where $\omega^{\mu}\omega_{\mu}=\omega_{\nu}g^{\nu\mu}\omega_{\mu}$,
while the  {local} outer product is given by 
\bea
\ket{\Phi}\bra{\Phi}&=&\bar{\phi}\phi\oplus \Big(\bar{\omega}_\mu{\omega}_{\nu} dx^{\mu}\otimes dx^{\nu}\Big)\nonumber \\
&&\oplus \bar{\zeta}_{\mu\nu}\zeta_{\rho\sigma}(dx^{\mu}\wedge dx^{\nu})\otimes(dx^{\rho}\wedge dx^{\sigma}).
\eea
Indicating here with ${d}$ the differential operator and with ${\delta}$ the codifferential operator, we define the Dirac operator $D$ as  $D={\delta}+{d}$, that is restricted to the space of  topological bosons. Thus we define the action of $D$ over $\ket{\Phi}$ as
\bea
{D}\ket{\Phi}&=&
-\nabla^{\mu}\omega_{\mu}\oplus (\nabla_{\mu}\phi -\nabla^{\rho} \zeta_{\rho\mu}) dx^{\mu}\nonumber \\
&&\oplus\nabla_{\mu}\omega_{\nu} dx^{\mu}\wedge dx^{\nu}.
\label{Dirac_K}
\eea
The metric $\tilde{\bf G}$ induced by the topological matter field will have  structure  similar to $\tilde{g}$ that we can write in full generality as
\bea
\tilde{\bf G}&=&G_{(0)}\oplus[G_{(1)}]_{\mu\nu}dx^{\mu}\otimes dx^{\nu}\nonumber\\
&&\oplus [G_{(2)}]_{\mu\nu\rho\sigma}(dx^{\mu}\wedge dx^{\nu})\otimes(dx^{\rho}\wedge dx^{\sigma}).
\eea
Proceeding as in the warm-up scenario we might wish to define the metric induced by the topological field as 
\bea
\tilde{\bf G}&=&\tilde{g}+\alpha\Big({D}\ket{\Phi}\bra{\Phi}D\Big).
\label{Ggen}
\eea
 {Thus in this general scenario we consider the two covariant topological metrics $\tilde{g}$ and $\tilde{\bf G}$ each given by the direct sum of metrics between  {scalars, vectors and bivectors.} As a reference on our notation for this general scenario we refer the reader to the Table \ref{table_general_scenario_I} and the Table \ref{table_general_scenario_II}.}\\
In our interpretation of the metric as a quantum density matrix, this would correspond to density matrix corresponding to a pure state $D\ket{\Phi}$. 
%where 
%\bea
%{D}\ket{\Phi}\bra{\Phi}D=\nabla^{\mu}\bar{\omega}_{\mu}\nabla^{\nu}\omega_{\nu}\oplus \nabla_{\mu}\bar{\phi}\nabla_{\nu}\phi \ dx^{\mu}\otimes dx^{\nu}.
%\eea
However, if the induced metric $\tilde{\bf G}$ is interpreted as a density matrix, it is natural to add further terms in the metric $\tilde{\bf G}$. These terms will depend on the topological field and the geometry of the space which will allow us to describe mixed states.
First,  we introduce in $\tilde{\bf G}$ a term proportional to $\ket{\Phi}\bra{\Phi}$. Specifically we introduce a term  $(m^2+\xi R)\ket{\Phi}\bra{\Phi}$ where $R$ is the Ricci scalar and $\xi$ might include the case of conformal coupling $\xi=(d-2)/(4(d-1))$.  This term can be also interpreted as a projector. Secondly, we introduce a term depending explicitly on the curvature of the manifold.
Since $\tilde{\bf G}$ involves the  metric for  {scalars, vectors and bivectors} on equal footing,  it is natural to consider a further term involving  $\tilde{\bm{\mathcal{R}}}$ given by the direct sum of the Ricci scalar $R$, the Ricci tensor of elements $R_{\mu\nu}$  and the Riemann tensor of elements $R_{\mu\nu\rho\sigma}$, i.e.
\bea
\tilde{\bm{\mathcal{R}}}&=&R\oplus \Big(R_{\mu\nu} dx^{\mu}\otimes dx^{\nu}\Big)\nonumber \\&&\oplus R_{\mu\nu\rho\sigma}(dx^{\mu}\wedge dx^{\nu})\otimes(dx^{\rho}\wedge dx^{\sigma}).
\eea
 Including $\tilde{\bm{\mathcal{R}}}$ into the metric $\tilde{\bf G}$ will allow as to describe more general metric matrices that are not decomposable into sum of projectors.
 
From these considerations, it follows that  in the induced metric $\tilde{\bf G}$,  we will substitute the term $D\ket{\Phi}\bra{\Phi}D$  with $\tilde{\bf M}$  defined as 
\bea
\tilde{\bf M}={D}\ket{\Phi}\bra{\Phi}D+(m^2+\xi R)\ket{\Phi}\bra{\Phi}.
\label{tildeT}
\eea
As we will see in the following paragraph this choice will allow us to  effectively overcome the first limitation of the warm-up scenario and to recover the Klein-Gordon equation in curved spacetime in full.
Furthermore we  consider also the additional term proportional to $\tilde{\bm{\mathcal{R}}}$ 
 and we  postulate that the metric $\tilde{\bf G}$ induced by the geometry and the matter fields is given by 
\bea
\tilde{\bf G}&=&\tilde{g}+\alpha{\tilde{\bf M}}-\beta{\tilde{\bm{\mathcal{R}}}}.
\label{tildeG}
\eea
As we will see in the following the addition of the term proportional to $\tilde{\bm{\mathcal{R}}}$ will allow  also to overcome the second limitation of the warm-up scenario.
Note that here $\alpha, \beta$ are positive constants. In particular, since we require $\tilde{\bf G}$ to be adimensional, we need to consider $\alpha=\alpha'\ell_P^d$ and $\beta=\beta'\ell_P^2$ where $\ell_P$ is the Planck length and $\alpha',\beta'$ are adimensional in the units $\hbar=c=1$. 

 {In the main body of this paper we will investigate only (bosonic) Dirac-K\"ahler matter fields. However gauge fields and fermionic Dirac fields can be included as well. For a discussion of the inclusion of  Abelian gauge field see  Appendix \ref{Ap_gauge}.}

Possibly  this approach could be extended to include also higher-forms. However, for simplicity, we consider here only topological matter fields formed by the direct sum between a 0-form, a 1-form and a 2-form as  this the minimal choice that will allow us   to include in the action the Ricci scalar, the Ricci and  the Riemann tensor explicitly.

\begin{figure}[!htb!]
  \includegraphics[width=\columnwidth]{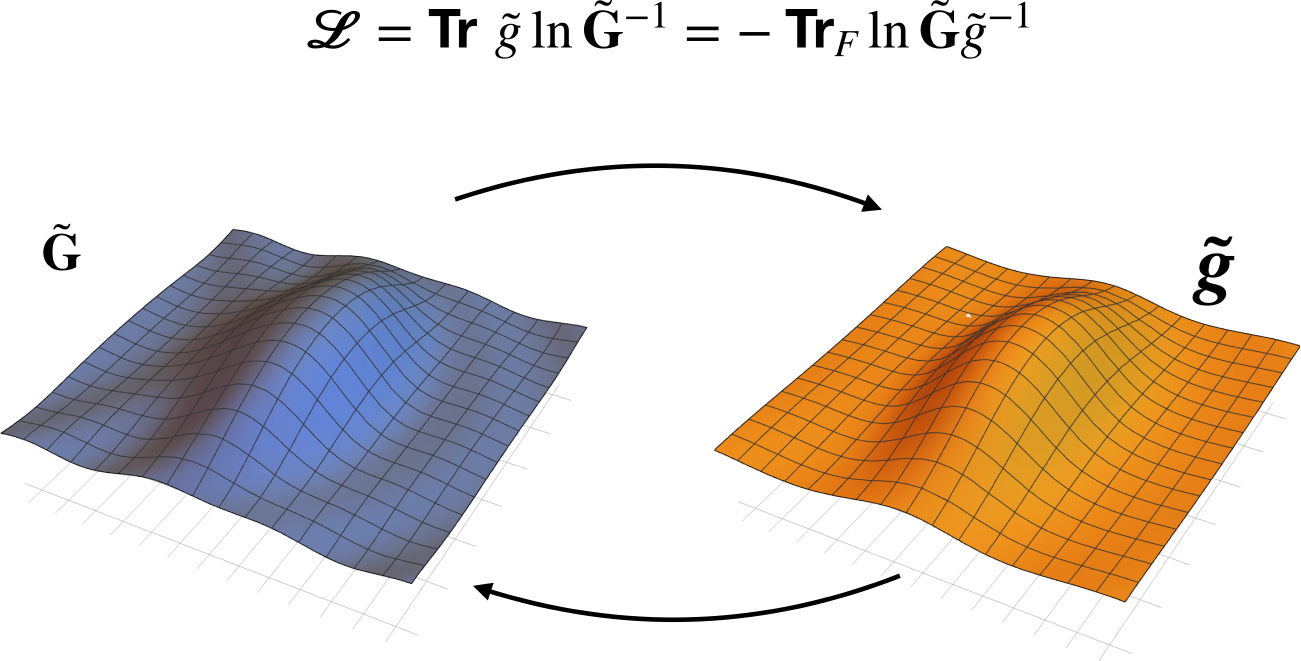}
  \caption{Schematic representation of this theoretical framework.The metric induced by the matter field $\tilde{\bf G}$ affects the metric of the manifold $\tilde{g}$ and vice versa the metric of the manifold affects the metric induced by the matter field. The considered Lagrangian is given by  the quantum relative entropy between the topological metric $\tilde{g}$ of spacetime and the topological metric induced by the topological matter fields $\tilde{\bf G}$. Since, by definition,  the entropy of the metric vanishes identically, the quantum relative entropy in the Lagrangian $\mathcal{L}$ reduces to a single term:  the quantum cross-entropy.}
  \label{fig1}
 \end{figure}

\subsection{Entropic topological and geometrical action}
We propose a statistical mechanics action formulated in terms of the quantum relative entropy between the metric $\tilde{g}$ and the metric $\tilde{\bf G}$ induced by the topological matter fields. In order to  define our action we need to extend the notion of eigenvalues to metric tensors  between  {bivectors.} This will allow to define the entropy and the quantum relative entropy in this novel framework. For a detailed discussion of this mathematical background see Appendix $\ref{ApA}$ that constitutes also the reference for our notation.
First of all we observe that 
the entropy associated to the metric $\tilde{g}$ remains zero. Indeed we define $\tilde{H}$ as  
\bea
\tilde{H}&=&\mbox{Tr}\tilde{g}\ln {\tilde{g}^{-1}}\nonumber \\&=&1\ln 1+\mbox{Tr} g\ln g^{-1}+\mbox{Tr} g_{(2)} \ln g_{(2)}^{-1}=0,
\eea
where we have used   Eq.(\ref{H}) and the identity derived in the Appendix $\ref{ApA}$ (in Eq.(\ref{gntrace})), 
\bea
\mbox{Tr} g_{(2)} \ln g_{(2)}^{-1}=0,
\eea
This equation is a direct consequence of the fact that not only $g$ but also $g_{(2)}$, has all the eigenvalues equal to one.

We are now in the position to consider the Lagrangian  given by the quantum relative entropy between $\tilde{g}$ and $\tilde{\bf G}$, 
\bea
\mathcal{L}&:=&-{\mbox{Tr}}\tilde{g}\ln {\tilde{ g}^{-1}}+{\mbox{Tr}}\tilde{g}\ln {\tilde{\bf G}^{-1}}
\label{Lt0}
\eea
 (see Figure \ref{fig1} for an illustration of the physical model beyond the choice of this Lagrangian).
 Since we have $\tilde{H}=0$ we obtain
\bea
\mathcal{L}&=&{\mbox{Tr}}\tilde{g}\ln {\tilde{\bf G}^{-1}}.
\label{Lt}
\eea
By treating separately the contributions of ${G_{(0)}}$  ${\bf G}_{(1)}$ and ${\bf G}_{(2)}$ we get the explicit expression for the Lagrangian $\mathcal{L}$ given by
\bea
\mathcal{L}&:=&\ln [{G_{(0)}}]^{-1}+\mbox{Tr} g\ln [{\bf G}_{(1)}]^{-1}\nonumber\\
&&+\mbox{Tr} g_{(2)}\ln [{\bf G}_{(2)}]^{-1}.\eea
Since the metric matrices $g_{(n)}$ have all their eigenvalues equal to one, using the notation developed in Appendix \ref{ApA} we can express this Lagrangian as well as 
\bea
\mathcal{L}&:=&-\mbox{Tr}_F\ln \ {\tilde{\bf G}}\tilde{g}^{-1}\nonumber \\&=&-\ln [{G_{(0)}}]-\mbox{Tr}_F \ln [{\bf G}_{(1)}]g^{-1}\nonumber \\&&-\mbox{Tr}_F \ln [{\bf G}_{(2)}]^{-1}{[g_{(2)}]}^{-1}.
\eea

The resulting statistical mechanics  action $\mathcal{S}$ associated to the Lagrangian $\mathcal{L}$  {is strictly related to the Araki quantum relative entropy~\cite{araki1975relative,ohya2004quantum,witten2018aps}, and can be formally derived from a mathematical theory of quantum operators  (see Appendix \ref{quantum})}. Specifically, the considered action  is given by 
\bea
\mathcal{S}=\frac{1}{\ell_P^d}\int \sqrt{|-{g}|}\mathcal{L} d{\bf r}.\eea
This action defines a modified theory of gravity. In the linearised limit  $\alpha'\ll 1$, $\beta'\ll 1$  this action reduces  to the Einstein-Hilbert action with zero cosmological constant ~\cite{einstein1915feldgleichungen,carroll2019spacetime}   coupled with the scalar topological field.  Indeed in this limit $\mathcal{L}$ reduces to
\bea
\mathcal{L}&=&3{\beta}R-\alpha  \bra{\Phi}{D}\tilde{g}^{-1}D\ket{\Phi}\nonumber \\&&-\alpha (m^2+\xi R)  \Big(|\phi|^2+\bar{\omega}^{\mu}\omega_{\mu}+\bar{\zeta}^{\mu\nu}\zeta_{\mu\nu}\Big).
\label{EH}
\eea
where
\bea
\bra{\Phi}{D}\tilde{g}^{-1}{D}\ket{\Phi}&=&|\nabla \phi |^2+|\nabla^{\mu}\omega_{\mu}|^2+\Big|\nabla^{\rho}{\zeta}_{\rho\mu}\Big|^2\nonumber \\
&&\hspace{-8mm}+\Big|\epsilon^{\mu\nu\rho}\nabla_{\mu}{\omega}_{\nu}\Big|^2.
\eea 
%\bea
%\bra{\Phi}{D}^2\ket{\Phi}&=&|\nabla &\phi |^2+|\nabla^{\mu}\omega_{\mu}|^2+\Big(\nabla_{\rho}&\bar\omega^{\rho\mu}\nabla^{\sigma}\bar{\omega}_{\sigma\mu}\Big)\nonumber \\
%&&\hspace{-8mm}+\frac{1}{4}\Big( (\nabla^{\mu}\bar{\omega}^{\nu}-\nabla^{\nu}\bar\omega^{\mu})(\nabla_{\mu}{\omega}_{\nu}-\nabla_{\nu}\omega_{\mu})\Big)
%\eea 
In the interesting limit in which $\omega_{\mu}=0$ and $\zeta_{\mu\nu}=0$ and the topological matter field is only scalar, Eq.(\ref{EH}) reduces  to the widely studied Einstein-Hilbert action with zero cosmological constant coupled with the scalar field \cite{mukhanov2007introduction},
\bea
\mathcal{L}&=&3{\beta}R-\alpha |\nabla \phi |^2-\alpha (m^2+\xi R)  |\phi|^2.
\label{EH2}
\eea
As already anticipated, this implies, among the other things, that this framework  overcomes the first and the second limitations of the warm-up scenario. In fact in this way  we recover the Klein-Gordon Lagrangian in curved spacetime in full.
Moreover, by  adding to $\tilde{\bf G}$ the term proportional to the curvature $\tilde{\bm{\mathcal{R}}}$ we solve the second limitation of the warm-up scenario, and we obtain that the entropic action allows for the full determination of the metric in the vacuum.

\subsection{Introduction of G-field $\tilde{\mathcal{G}}$ and the  field $\tilde{\Theta}$ }
In this section our goal is  to investigate the property of the modified gravity emerging from the entropic action. To this end we introduce the auxiliary G-field $\tilde{\bm{\mathcal{G}}}$ and the auxiliary field $\tilde{\bm{\Theta}}$. In this way we will be able to show in the following that the proposed entropic action can be interpreted as the sum of a dressed Einstein-Hilbert action with an emergent positive cosmological constant only depending on the G-field $\tilde{\bm{\mathcal{G}}}$ and a matter action.

As mentioned before  we can express the Lagrangian $\mathcal{L}$ given by Eq.(\ref{Lt}) as 
\bea
\mathcal{L}=-\mbox{Tr}_F\ln \ {\tilde{\bf G}}\tilde{g}^{-1}.
\label{L2}
\eea
This Lagragian is nonlinear in $\tilde{\bf G}$ and thus in $\tilde{\bm{\mathcal{R}}}$.

Several modified gravity actions are questioned because they give rise to  theories
with  derivatives of the metric higher than two. Such theories can be affected by the Ostrogradsky instability,  however non-linear theories not suffering from this pathology are also known, most notably the $f(R)$ theories \cite{woodard2007avoiding,sotiriou2010f}.
Thus an important question is whether the proposed theory is also affected by the Ostrogradsky instability or rather, it generalizes and  extends the realm of viable modified gravity theories beyond the $f(R)$ theories. In particular the Ostrogradsky instability can arise when the Lagrangian depends on derivatives of order higher than two of the fields, and thus, in the gravitational setting on higher powers of the curvature $\tilde{\mathcal{R}}$.

In order to tackle this question we observe that the Lagrangian $\mathcal{L}$ has  an easy expression as it is only dependent on the product ${\tilde{\bf G}}\tilde{g}^{-1}$.
Thus, by introducing an auxiliary field $\tilde{\bm{\Theta}}$ and imposing the constraints
\bea
\tilde{\bf G}\tilde{g}^{-1}=\tilde{\bm{\Theta}}.
\label{constraints}
\eea
 with Lagrangian multipliers $\tilde{\bm{\mathcal{G}}}$ (constituting another auxiliary field) that we will call the {\em  G-field}, we can reduce our theory to a theory driven by the Lagrangian $\tilde{\mathcal{L}}$ given by 
\bea
\tilde{\mathcal{L}}=-\mbox{Tr}_F\ln \tilde{\bm{\Theta}}-\mbox{Tr}_{F}\tilde{\bm{\mathcal{G}}}(\tilde{\bf G}\tilde{g}^{-1}-\tilde{\bm\Theta}).
\label{LW}
\eea
This Lagrangian  is now linear in $\tilde{\bf G}\tilde{g}^{-1}$.  {Note that a conservative interpretation of this transformation will not give a physical interpretation to the auxiliary fields. Thus, according to this point of view, the auxiliary fields are not changing the physics of the original Lagrangian and this theory might be prone to Ostrogradsky instability. However, we known from statistical physics that Lagrangian multipliers can also acquire a rather physical meaning. For instance the temperature and the chemical potential are Lagrangian multipliers that are measurable. Here we would like to embrace this point of view and give a physical meaning to the introduced Lagrangian multipliers as physical fields defined on our manifold $\mathcal{K}$. Interpreting in this way the Lagrangian multipliers  changes the  physics of the problem as the phase space associated to the equations of motion acquire new dimensions and  the fixed points as well as the  initial value problem will involve the newly introduced fields.  In this latter interpretation, introducing the auxiliary fields and giving a physical meaning to them as encoding for new fields, we have reduced our theory to a theory linear in $\tilde{\bm{\mathcal{R}}}$. In particular, as we will see below, the equations of motion will involve at most second order derivatives of the fields, thus the model { might} avoid the Ostrogradsky instability. {  A definitive answer to
the stability question would follow from further analysis (for example a
Hamiltonian analysis).}

The  fields $\tilde{\bm{\Theta}}$ and $\tilde{\bm{\mathcal{G}}}$ are given by
\bea
\tilde{\bm{\Theta}}&=&{\Theta}_{(0)}\oplus {\bm{\Theta}}_{(1)}\oplus {\bm{\Theta}}_{(2)},\nonumber \\
\tilde{\bm{\mathcal{G}}}&=&{\mathcal{G}}_{(0)}\oplus {\bm{{\mathcal{G}}}}_{(1)}\oplus {\bm{{\mathcal{G}}}}_{(2)},\eea
with  ${\bm{{\Theta}}}_{(1)}$ having elements ${[\bm{\Theta}_{(1)}]}_{\mu}^{\ \nu}$, ${\bm{\Theta}}_{(2)}$ having elements ${[\bm{\Theta}_{(2)}]}_{\mu\nu}^{\ \ \rho\sigma}$ 
and 
${\bm{{\mathcal{G}}}}_{(1)}$ having elements ${[\bm{\mathcal{G}}_{(1)}]}_{\mu}^{\ \nu}$, ${\bm{\mathcal{G}}}_{(2)}$ having elements ${[\bm{\mathcal{G}}_{(2)}]}_{\mu\nu}^{\ \ \rho\sigma}$.
The term $\mathcal{W}$, of the Lagrangian $\tilde{\mathcal{L}}$ that is  linear in $\tilde{\bf G}\tilde{g}^{-1}$, can be expressed  as
\bea
\mathcal{W}=\mbox{Tr}_{F}\tilde{\bm{\mathcal{G}}}(\tilde{\bf G}\tilde{g}^{-1}-\tilde{\bm\Theta})=\sum_{n=0}^2\mathcal{W}_n,
\eea
with
\bea
\hspace{-5mm}\mathcal{W}_0&=&{\mathcal{G}}_{(0)}(G_{(0)}-{\Theta}_{(0)}),\nonumber \\
\hspace{-5mm}\mathcal{W}_1&=&{[{\mathcal{G}}_{(1)}]}_{\rho}^{\ \mu}\Big({[{G_{(1)}}]}_{\mu\nu}g^{\nu\rho}-{[{\Theta}_{(1)}]}_{\mu}^{\ \rho}\Big),\nonumber \\
\hspace{-5mm}\mathcal{W}_2&=&{[{\mathcal{G}}_{(2)}]}_{\eta\theta}^{\ \ \mu\nu}\Big({[{G_{(2)}}]}_{\mu\nu\rho\sigma}{[g_{(2)}]}^{\rho\sigma\eta\theta}-{[{\Theta}_{(2)}]}_{\mu\nu}^{\ \ \eta\theta}\Big).
\label{W}
\eea
Note that the use Lagrange multipliers in the present  theory extends the equivalence between the $f(R)$ theories and Brans-Dicke Theory achieved through a Legendre transformation \cite{woodard2007avoiding,sotiriou2010f}. However using an extensive number of Lagrange multipliers might be considered a dangerous mathematical operation.
For the sake of simplicity here we work under the assumption that $\tilde{\mathcal{L}}$ is equivalent to ${\mathcal{L}}$.
Whereas it might be proven that $\mathcal{L}$ is not equivalent to $\tilde{\mathcal{L}}$ we assume that our true theory is given by $\tilde{\mathcal{L}}$ which we might always consider as a ``canonical" version of $\mathcal{L}$.

The resulting statistical mechanics  action $\tilde{\mathcal{S}}$ associated to the Lagrangian $\tilde{\mathcal{L}}$ is  given by 
\bea
\tilde{\mathcal{S}}=\frac{1}{\ell_P^d}\int \sqrt{|-{g}|}\tilde{\mathcal{L}} d{\bf r}.
\label{hS}\eea
We are now in the position to derive the equations for the matter fields $\Phi$, the metric $g_{\mu\nu}$ and  the fields $\tilde{\bm{\Theta}}$ and $\tilde{\bm{\mathcal{G}}}$ by considering the variation of $\tilde{\mathcal{L}}$ with respect to these fields.
\subsection{Equation of motion for the matter fields}
\label{sec:matter}
Here we consider the action $\tilde{\mathcal{L}}$ defined in Eq.(\ref{LW})  depending linearly on  $\tilde{\bf G}$ which is encoding for the  matter fields. By using the definition of $\tilde{\bf G}$ given by Eq.(\ref{tildeG}), we can derive the equation of motion of the matter fields by considering the variation with respect to $\bra{\Phi}$, getting,
\bea
{D}{\tilde{g}_{\mathcal{G}}}^{-1}{D}\ket{\Phi}+{\tilde{g}_{\mathcal{G}}}^{-1}(m^2+\xi R)\ket{\Phi}=0.
\label{matter_eq}
\eea
Here   $\tilde{g}_{\mathcal{G}}$, given by 
 \bea
 \tilde{g}_{\mathcal{G}}=\tilde{\bm{\mathcal{G}}}^{-1}g,
\label{gG} 
 \eea
can be interpreted as the  {\em dressed metric} which affects the matter fields.
Specifically, $g$ can be seen as a bare metric and the G-field $\tilde{\bm{\mathcal{G}}}^{-1}$ can be seen as a  dressing of this metric that gives rise to the dressed metric $\tilde{g}_{\mathcal{G}}$ given by Eq.(\ref{gG}).

\subsection{Modified gravity}
We now turn to the equations for modified gravity considering variation of our action $\tilde{\mathcal{S}}$ with respect to the metric $g$ and to the  fields $\tilde{\bm\Theta}$ and $\tilde{\bm{\mathcal{G}}}$.

\subsubsection{Variation with respect to $\tilde{\bm{\mathcal{G}}}$}
The variation with respect to  the fields $\tilde{\bm{\mathcal{G}}}$ enforces the constraints in Eq.$(\ref{constraints})$ which, by using the expression for $\tilde{\bf G}$  given by Eq.(\ref{tildeG}), can be expressed explicitly  as 
\bea
\tilde{\bm\Theta}=\tilde{\bf I}+\alpha\tilde{\bf M}\tilde{g}^{-1}-\beta\tilde{\bm{\mathcal{R}}}\tilde{g}^{-1},
\label{Thetaeq}
\eea
where $\tilde{\bf I}$ is the (topological) identity.
Since $\alpha=\alpha'\ell_P^4$ and $\beta=\beta'\ell_P^2$  in the regime of low energies and low curvatures, we have that $\tilde{\bm\Theta}$ is a small perturbation of the identity $\tilde{\bf I}$.

Note that these are the equations that determine the relation between  the fields $\tilde{\bm{\Theta}}$ the curvature $\tilde{\bm{\mathcal{R}}}$ and the matter field determining $\tilde{\bf M}$.
We can write these equations by separating the contribution corresponding to the metric for $0$-forms, $1$-forms and $2$-forms obtaining  
\bea
\Theta_{(0)}&=&1+\alpha M_{(0)}-\beta R,\nonumber \\
{[\Theta_{(1)}]}_{\mu}^{\ \nu}&=&\delta_{\mu}^{\ \nu}+\alpha{[{M}_{(1)}]}_{\mu\rho}g^{\rho\nu}-\beta R_{\mu\rho}g^{\rho\nu},\nonumber \\
{[\Theta_{(2)}]}_{\vec{\mu}}^{\ \vec{\nu}}
&=&\delta_{\vec{\mu}}^{\ \vec{\rho}}+\alpha{[{M}_{(2)}]}_{\vec{\mu}\vec{\rho}}{[g_{(2)}]}^{\vec{\rho}\vec{\nu}}-\beta R_{\vec{\mu}\vec{\rho}}{[g_{(2)}]}^{\vec{\rho}\vec{\nu}},\nonumber 
\eea
where we have indicated a pair of indices with a vector symbol, e.g.  $\vec{\mu}$.
Since the trace of the Ricci tensor as well as the trace of the Riemann tensor, are both equal to the Ricci scalar, upon performing the trace of these expressions, we get some constraints on the trace of the  fields ${\bm\Theta}_{(n)}$, i.e.
\bea
\begin{pmatrix}d\\n\end{pmatrix}-\mbox{Tr}_F {\bm\Theta}_{(n)}+\alpha\mbox{Tr}_F {\bf M}_{(n)}g^{-1}_{(n)}=\beta R,
\eea
for any $n\in \{0,1,2\}$.

\subsubsection{Variation with respect to $\tilde{\bm{\Theta}}$}
Thanks to the logarithmic nonlinearity in $\tilde{\mathcal{L}}$ the variation of the action $\tilde{\mathcal{S}}$ with respect to ${\tilde{\bm{\Theta}}}$ simply reads 
\bea
\tilde{\bm{\Theta}}^{-1}=\tilde{\bm{\mathcal{G}}}.\label{inv}
\eea
Thus we obtain that the   field $\tilde{\bm{\Theta}}$ corresponds to the inverse of the dressing G-field $\tilde{\bm{\mathcal{G}}}$.
Using Eqs.(\ref{Thetaeq}) we  find that the equation of motion for the G-field $\tilde{\bm{\mathcal{G}}}$  is given by
\bea
\tilde{\bm{\mathcal{G}}}^{-1}=\tilde{\bf I}+\alpha\tilde{\bf M}\tilde{g}^{-1}-\beta\tilde{\bm{\mathcal{R}}}\tilde{g}^{-1},
\label{Gm1b}
\eea
and we can simply eliminate the   field $\tilde{\bm\Theta}$,  from the action.
In this way  the considered action $\tilde{\mathcal{S}}$ can be decomposed into two terms as
\bea
\tilde{\mathcal{S}}=\beta\mathcal{S}_G+\alpha\mathcal{S}_M,
\label{St}
\eea
with 
\bea
\hspace{-8mm}\mathcal{S}_G=\frac{1}{\ell_P^d}\int \sqrt{|-g|}\mathcal{L}_G d{\bf r},\quad
\mathcal{S}_M=\frac{1}{\ell_P^d}\int \sqrt{|-g|} \mathcal{L}_M d{\bf r}.\nonumber \\
\label{SEt}
\eea
The Lagrangians $\mathcal{L}_G$ and $\mathcal{L}_M$ appearing in the above definition of  $\mathcal{S}_G$ and $\mathcal{S}_M$ can be expressed as  
\bea
\mathcal{L}_G= \Big(\mathcal{R}_{\mathcal{G}}-2\Lambda_{\mathcal{G}}\Big),\quad
\mathcal{L}_M=-{\mathcal{M}}_{\mathcal{G}}.
\label{LEt}
\eea
where $\mathcal{R}_{\mathcal{G}}$ is the {\em dressed Ricci scalar} and $\mathcal{M}_{\mathcal{G}}$ includes all dependence with the matter fields $\ket{\Phi}$ while $\Lambda_{\mathcal{G}}$ is the emergent  positive cosmological constant. Specifically we have
\bea
\mathcal{R}_{\mathcal{G}}=\mbox{Tr}_{F}\tilde{g}_{\mathcal{G}}^{-1}\tilde{\bm{\mathcal{R}}},\quad 
\mathcal{M}_{\mathcal{G}}= \mbox{Tr}_{F}\tilde{g}_{\mathcal{G}}^{-1}\tilde{\bm{M}},\nonumber \\
\hspace{-5mm}\Lambda_{\mathcal{G}}=\frac{1}{2\beta}\mbox{Tr}_{F}\Big(\tilde{\bm{\mathcal{G}}}-\tilde{\bf I}-\ln \tilde{\bm{\mathcal{G}}}\Big).
\label{dressed_eq}
\eea
From this reformulation of the action $\tilde{S}$ we see that the Lagrangian $\mathcal{L}_G$  has a the structure  of a {\em dressed Einstein-Hilbert action} and depends on the metric and the  G-field $\tilde{\bm{\mathcal{G}}}$. In particular the Ricci scalar $R$ is substituted to the dressed Ricci scalar $\mathcal{R}_{\mathcal{G}}$ and the role of the cosmological constant is played by $\Lambda_{\mathcal{G}}$ that is non-negative and only dependent on the G-field   $\tilde{\bm{\mathcal{G}}}$. Whenever this field is  close to the identity, e.g. $\tilde{\bm{\mathcal{G}}}\simeq \tilde{\bf I}+\tilde{\bm\epsilon}$ we obtain that $\Lambda_{\mathcal{G}}$ is positive but very small, i.e. its first non-trivial term is quadratic in $\tilde{\bm\epsilon}$. Thus the emergent positive cosmological constant $\Lambda_{\mathcal{G}}$ is small in this limit and only determined by the G-field. 
Finally we recover as already noticed, that the action $\tilde{S}$ reduces to the Einstein-Hilbert with zero cosmological constant in the low coupling limit.

\subsubsection{Variation with respect to ${g}$}
The modified Einstein equations are obtained by performing the variation of the action $\tilde{\mathcal{S}}$ with respect to $g$.
Let us define the stress-energy tensor $\bm{\mathcal{T}}$ in the usual way with elements ${\mathcal{T}}_{\mu\nu}$ given by  
\bea
-\frac{1}{\sqrt{|-g|}}\frac{\delta \mathcal{S}_M}{dg_{\mu\nu}}={\mathcal{T}}_{\mu\nu}.
\eea
With this notation we can expressed the modified Einstein equations as 
\bea
{R}^{\mathcal{G}}_{(\mu\nu)}-\frac{1}{2}{g}_{\mu\nu}\Big(\mathcal{R}_{\mathcal{G}}-2\Lambda_{\mathcal{G}}\Big)+{{\mathcal{D}}}_{(\mu\nu)} =\kappa{{\mathcal{T}}_{(\mu\nu)}},
\label{modEin}
\eea
where $\kappa=\alpha/\beta$, $(\mu\nu)$ indicates symmetrization of the indices, ${R}^{\mathcal{G}}_{\mu\nu}$ are the elements or the {\em dressed Ricci tensor} given by
\bea
{R}^{\mathcal{G}}_{\mu\nu}&=&{\mathcal{G}_{(0)}}R_{\mu\nu}+{[{\mathcal{G}_{(1)}}]}_{\mu}^{\ \rho}R_{\rho\nu}-{[\mathcal{G}_{(2)}]}_{\rho_1\rho_2\mu\eta}R_{\nu}^{\ \eta\rho_1\rho_2}\nonumber \\
&&+2{[{\mathcal{G}_{(2)}}]}_{\mu}^{\ \eta\rho_1\rho_2}R_{\rho_1\rho_2\nu\eta},
\eea
and ${{\mathcal{D}}}_{\mu\nu}$ are the elements depending on second derivatives of the G-field $\tilde{\bm{\mathcal{G}}}$ given by 
\bea
{{\mathcal{D}}}_{\mu\nu}&=&(\nabla^{\rho}\nabla_{\rho}g_{\mu\nu}-\nabla_{\mu}\nabla_{\nu}){\mathcal{G}_{(0)}}-\nabla^{\rho}\nabla_{\nu}{[\mathcal{G}_{(1)}]}_{(\rho\mu)}
\nonumber \\
&&+\frac{1}{2}
\nabla^{\rho}\nabla_{\rho}{[\mathcal{G}_{(1)}]}_{\mu\nu}+\frac{1}{2}\nabla^{\rho}\nabla_{\eta}{[\mathcal{G}_{(1)}]}_{\rho\eta}g_{\mu\nu}\nonumber \\
&&+\nabla^{\eta}\nabla^{\rho}{[\mathcal{G}_{(2)}]}_{\mu\rho\nu\eta}+\nabla^{\rho}\nabla^{\eta}{[\mathcal{G}_{(2)}]}_{\eta\mu\rho\nu}\nonumber \\
&&+\frac{1}{2}[\nabla^{\rho},\nabla^{\eta}]{[\mathcal{G}_{(2)}]}_{\rho\eta\mu\nu}.
\eea
It follows that the modified Einstein equations involve only second derivatives of the metric and second derivatives of the field $\tilde{\bm{\mathcal{G}}}$. However these equations might have more solutions than the Einstein equations.  
A detailed study of the solutions and the viability of these equations is beyond the scope of this work and will be the subject of future investigations.
\subsubsection{Discussion}
In summary the introduction of the G-field turns the proposed entropy action Eq.(\ref{Lt}) into the action Eq.(\ref{St}) involving a dressed Einstein-Hilbert action and a dressed matter action given by Eq.({\ref{SEt}}) depending on the matter and the gravity dressed Lagrangians Eq.(\ref{LEt}) respectively. The gravitational dressed action $\mathcal{L}_{\mathcal{G}}$ displays an emergent positive cosmological constant $\Lambda_{\mathcal{G}}$ only dependent on the G-field and given by Eq.(\ref{dressed_eq}). The equations of motion associated to this action involve solving Eqs.(\ref{matter_eq}) for the matter fields, and Eqs. (\ref{Gm1b}) and Eqs. (\ref{modEin}) for the metric and the G-field. All these equations involve at most second derivatives of the fields.

\section{Conclusions}
\label{sec:conclusions}
This work proposes a modified theory of gravity emerging from statistical mechanics and information theory action. The fundamental idea of this theory is to associate the metric to a  {quantum operator, playing the role of a {\em renormalizable and effective density matrix}}. In particular two metrics are discussed: the metric of  spacetime that fully defines its geometry and the metric induced by matter-fields that are effectively curving the space.  The interplay between geometry of spacetime  and matter fields is explicitly captured by the proposed action given by the quantum relative entropy between the metric and the metric induced by the matter fields.

Here  a  Lorentzian theory  consistent with this fundamental physical interpretation of gravity is formulated. In order to do this we have built the necessary mathematical background to define the entropy and the cross-entropy of metric tensors in a Lorentzian well defined way. A modified gravity is emerging from this framework when the matter fields are described by topological  Dirac-K\"ahler bosons formed by the direct sum between a 0-form, a 1-form and a 2-form and the induced metric also depends on the curvature of spacetime. The modified Einstein equations reduce to the Einstein equations in the regime of low coupling.
By introducing the G-field we obtain  the modified Einstein equations and the equations of motion for the matter and the G-field. From this theory it emerges that the proposed entropic action takes the form of a sum between a  dressed Einstein-Hilbert action and a matter action. Interestingly, the dressed Einstein-Hilbert action displays an emergent positive cosmological constant that only depends on the G-field.
Moreover, thanks to the introduction of the G-field,   the  equations of modified gravity remains second order in the metric, in the matter and in the G-field.

 {The interpretation of the topological metrics tensors as a quantum operators, or effective density matrix  where we have relaxed the constraints of having  unitary trace, and we have required the existence of the inverse is shown to be very useful}. These choices are motivated here by the necessity to have a Lorentzian invariant theory. 
%Moreover this generalization of the notion of density matrices points out to   connections . 
 {Here we have established the relation of the adopted quantum relative action with Araki quantum relative entropy \cite{araki1975relative}, opening the perspective of using the theory of quantum von Neumann algebras and the theory of entanglement~\cite{witten2018aps} to investigate further the properties of the proposed theory.}
 
In conclusion, we hope that this approach can help identify the deep connections between gravity,  {quantum mechanics} and statistical physics. Given the interpretation of the metrics as a  {quantum operators} our hope is that this approach will also be instrumental to formulate  approaches to quantum gravity in second quantization. Finally future investigation might explore the role of the  the G-field in  dark matter. Future directions in this line of  research involve  also the investigation of the proposed entropic action under the renormalization group,  and possible connections with phenomenology and experimental results. \\

This work was partially supported by a grant from the Simons Foundation. 
The author would like to thank the Isaac Newton Institute for Mathematical Sciences, Cambridge, for support and hospitality during the programme Hypergraphs: Theory and Applications, where work on this paper was undertaken. This work was supported by EPSRC grant EP/V521929/1. 
\appendix

\section{Inclusion of an Abelian gauge field}
\label{Ap_gauge}

{  An Abelian gauge field $A_{\mu}$ can be easily included in the general framework introduced in this work.
We consider the operator $\tilde{\bf F}$ defined as 
\bea
\tilde{\bf F}=0\oplus 0^{\mu}dx_{\mu}\oplus F_{\mu\nu}F_{\rho\sigma}\Big(dx^{\mu}\wedge dx^{\nu}\Big)\otimes\Big(dx^{\rho}\wedge dx^{\sigma}\Big),\nonumber 
\eea
where $F_{\mu\nu}=\nabla_{\mu}A_{\nu}-\nabla_{\nu}A_{\mu}$
and we consider the following expression for the metric induced by matter fields
\bea
\tilde{\bf G}&=&\tilde{g}+\alpha\Big({\tilde{\bf M}+\tilde{\bf F}}\Big)-\beta{\tilde{\bm{\mathcal{R}}}}.
\label{tildeG}
\eea
Note that in this expression we modify as well the definition of $\tilde{\bf M}$ as the Dirac operator will be defined as in Eq.(\ref{Dirac_K}) by performing the minimal substitution in the covariant derivative
\bea
\nabla_{\mu}\to \nabla_{\mu}^{(A)}=\nabla_{\mu}-\textrm{i}eA_{\mu}.
\eea}

\section{Eigenvalues and entropy for metrics ${\bf G}_{(1)}$ and for metrics ${\bf G}_{(n)}$}
\label{ApA}
In this Appendix our goal is to use algebraic geometry
\cite{renteln2014manifolds} to 
 define the eigenvalues of the metrics ${\bf G}_{(n)}$ considered as quantum operators between $n$-forms and their associated entropy.
In order to establish the foundation of this treatment let us revisit the treatment outlined in Sec. \ref{sec:preliminaries} for  defining the eigenvectors  and the associated entropy of the metrics ${\bf G}_{(1)}$ viewed as quantum operators between $1$-forms and treat on the same footing metrics tensor ${\bf G}_{(n)}$ viewed as quantum operators between  $n$-forms with arbitrary values of $n$. 
\subsection{Eigenvalues of metric matrices  ${\bf G}_{(n)}$}

A one form can be written as 
\bea
\omega=\omega_{\rho} dx^{\rho}.
\eea
The metric ${\bf G}_{(1)}$ can be interpreted as a quantum operator  among two of such $1$-forms and can be written as 
\bea
{\bf G}_{(1)}={[G_{(1)}]}_{\mu\nu}dx^{\mu}\otimes dx^{\nu},
\eea
where ${[G_{(1)}]}_{\mu\nu}$ is Hermitian.
The $\cdot$ product between the metric ${\bf G}_{(1)}$ and the generic one form $\omega$ is defined as 
\bea
{\bf G}_{(1)}\cdot \omega=\hat{\omega}=\hat{\omega}_{\mu}dx^{\mu},
\label{cdot1}
\eea
where, by definition
\bea
\hat{\omega}_{\mu}:={[G_{(1)}]}_{\mu\nu}\omega_{\rho}\Avg{dx^{\nu},dx^{\rho}},
\label{1s}
\eea
where here and in the following $\Avg{,}$ indicates the inner product between  $n$-forms. Thus for canonical $1$-forms we have \bea
\Avg{dx^{\nu},dx^{\rho}}=g^{\nu\rho}.
\eea 
Inserting this expression in Eq.(\ref{1s}) we get,
\bea
\hat{\omega}_{\mu}={[G_{(1)}]}_{\mu\nu}g^{\nu\rho}\omega_{\rho}.
\eea
Summarizing we have shown that  according to our definition we have
\bea
{\bf G}_{(1)}\cdot \omega={[G_{(1)}]}_{\mu\nu}g^{\nu\rho}\omega_{\rho} dx^{\mu}.
\eea
The eigenvalue problem of a metric ${\bf G}_{(1)}$  is then defined as 
  \bea
{\bf G}_{(1)}\cdot \omega\equiv\lambda{\omega},\eea
or equivalently in matrix form as 
\bea
{[G_{(1)}]}_{\mu\nu}g^{\nu\rho}\omega_{\rho}=\lambda\omega_{\mu}.
\eea
 {Note that ${[{\bf G}_{(1)}]}_{\mu\nu}$ admits also an interpretation as metric between vectors.} 
We are now in the position to extent the formalism to the metrics ${\bf G}_{(2)}$ considered as quantum operators among two  forms of order two. 
The generic two form $\zeta$ is defined as
\bea
\zeta=\zeta_{\eta\theta} dx^{\eta}\wedge dx^{\theta},
\eea
where $\zeta_{\eta\theta}=-\zeta_{\theta\eta}$, i.e. where $\zeta_{\eta\theta}$ is antisymmetric.

The metric ${\bf G}_{(2)}$ considered as a quantum operator between two two-forms is defined as 
\bea
{\bf G}_{(2)}=[G_{(2)}]_{\mu\nu\rho\sigma}(dx^{\mu}\wedge dx^{\nu})\otimes(dx^{\rho}\wedge dx^{\sigma}).
\eea
where ${\bf G}_{(2)}$ is antisymmetric in the first $2$ indices and in the last $2$ indices and Hermitian under the exchange of the first $2$ indices with the second $2$ indices.
Following a similar line of reasoning used to define the eigenvalue of the metric ${\bf G}_{(1)}$ we define the contraction
\bea
{\bf G}_{(2)}\cdot\zeta=\hat{\zeta}
=\hat{\zeta}_{\mu\nu} dx^{\mu}\wedge dx^{\nu},
\label{cdot2}
\eea
\ 
where, by definition $\hat{\zeta}_{\mu\nu}$ is defined as   
\bea
\hspace{-5mm}\hat{\zeta}_{\mu\nu}&:=&\frac{1}{2}
{[G_{(2)}]}_{\mu\nu\rho\sigma}\zeta_{\eta\theta}\Avg{dx^{\rho}\wedge dx^{\sigma},dx^{\eta}\wedge dx^{\theta}}.\eea
Performing the inner products we thus get
\bea
\hat{\zeta}_{\mu\nu}&=&\frac{1}{2}{[G_{(2)}]}_{\mu\nu\rho\sigma}\zeta_{\eta\theta}\Big(g^{\rho\eta}g^{\sigma\theta}-g^{\rho\theta}g^{\sigma\eta}\Big)\nonumber \\
&=&{[G_{(2)}]}_{\mu\nu\rho\sigma}{[g_{(2)}]}^{\rho\sigma\eta\theta}\zeta_{\eta\theta},
\eea
where we have used the definition of $g_{(2)}$ given in Eq.(\ref{g2}) that we rewrite here for convenience, 
\bea
{[g_{(2)}]}_{\mu\nu\rho\sigma}=\frac{1}{2}(g_{\mu\rho}g_{\nu\sigma}-g_{\mu\sigma}g_{\nu\rho}).
\eea
Summarizing we have shown that  according to our definition we have
\bea
{\bf G}_{(2)}\cdot\zeta={[G_{(2)}]}_{\mu\nu\rho\sigma}{[g_{(2)}]}^{\rho\sigma\eta\theta}\zeta_{\eta\theta} dx^{\mu}\wedge dx^{\nu}.
\eea
In total analogy with the previous case, the eigenvalue problem associated to the metric ${\bf G}_{(2)}$ viewed as quantum operator among two $2$-forms, is defined as 
  \bea
{\bf G}_{(2)}\cdot \zeta \equiv\lambda{\zeta},\eea
or equivalently,
\bea
{[G_{(2)}]}_{\mu\nu\rho\sigma}{[g_{(2)}]}^{\rho\sigma\eta\theta}\zeta_{\eta\theta}=\lambda\zeta_{\mu\nu}.
\label{eig2_b2}
\eea
 {Note that ${[{\bf G}_{(2)}]}_{\mu\nu\rho\sigma}$ admits also an interpretation as metric between bivectors.} 
Let us now consider the particular case in which ${\bf G}_{(2)}={g}_{(2)}$. In this case the eigenvalue problem Eq.(\ref{eig2_b2}) reads
\bea
\frac{1}{4}\Big(g_{\mu\rho}g_{\nu\sigma}-g_{\mu\sigma}g_{\nu\rho}\Big)\Big(g^{\rho\eta}g^{\sigma\theta}-g^{\rho\theta}g^{\sigma\eta}\Big)\zeta_{\eta\theta}=\lambda\zeta_{\mu\nu}.\nonumber 
\eea
which has solution 
\bea
\lambda=1,\eea
for any arbitrary $2$-form associated to the antisymmetric tensor $\zeta_{\mu\nu}$. Thus the degeneracy of the eigenvalue is given by  $d(d-1)/2$.
%Similarly it can be shown that the eigenvalues for the metric ${g}_{(n)}$ also remain identically equal to one for any $0\leq n\leq d$.

The treatment  of the eigenvalues associated to the  metrics tensors ${\bf G}_{(n)}$ considered as quantum operators between two $n$-forms can be performed in a straightforward way, following similar steps.
Let us observe that the generic $n$-form is given by 
\bea
\zeta=\zeta_{{\nu_1}\nu_2\ldots\nu_n}dx^{\nu_1}\wedge dx^{\nu_2},\wedge\ldots\wedge dx^{\nu_n}.
\eea
where $\zeta_{{\nu_1}\nu_2\ldots\nu_n}$ is antisymmetric,
while the generic expression for ${\bf G}_{(n)}$ is given by 
\begin{widetext}
\bea
G_{(n)}={[G_{(n)}]}_{\mu_1\mu_2\ldots\mu_n\nu_1\nu_2\ldots\nu_n}(dx^{\mu_1}\wedge dx^{\mu_2}\wedge\ldots dx^{\mu_n})\otimes(dx^{\nu_1}\wedge dx^{\nu_2},\wedge\ldots\wedge dx^{\nu_n}),\eea
\end{widetext}
where ${\bf G}_{(n)}$ is antisymmetric in the first $n$ indices and in the last $n$ indices and Hermitian under the exchange of the first $n$ indices with the second $n$ indices.

For instance, it is instructive to discuss explicitly the form of the metric $g_{(n)}$ induced on $n$-forms by $g$. This can be immediately shown to be  given by
\begin{widetext}
\bea
g_{(n)}=\prod_{i=1}^ng_{\mu_i\rho_i}(dx^{\mu_1}\wedge dx^{\mu_2}\wedge\ldots dx^{\mu_n})\otimes(dx^{\nu_1}\wedge dx^{\nu_2},\wedge\ldots\wedge dx^{\nu_n}),\eea
%\end{widetext}
or, more explicitly, using the antisymmetric properties of the $n$ forms, as
%\begin{widetext}
\bea
g_{(n)}={[g_{(n)}]}_{\mu_1\mu_2\ldots\mu_n\nu_1\nu_2\ldots\nu_n}(dx^{\mu_1}\wedge dx^{\mu_2}\wedge\ldots dx^{\mu_n})\otimes(dx^{\nu_1}\wedge dx^{\nu_2},\wedge\ldots\wedge dx^{\nu_n}),\eea
\end{widetext}
with 
\bea
{[g_{(n)}]}_{\mu_1\mu_2\ldots\mu_n\nu_1\nu_2\ldots\nu_n}=\frac{1}{n!}\delta_{\nu_1,\nu_2\ldots\nu_n}^{\rho_1\rho_2\ldots \rho_n}\prod_{i=1}^ng_{\mu_i\rho_i}.\eea
Here and in the following we will use the notation 
${[g_{(n)}]}_{\vec{\mu}\vec{\nu}}$ of indicating the element of this tensor where $\vec{\mu}=(\mu_1\mu_2\ldots \mu_n)$ and $\vec{\nu}=(\nu_1\nu_2\ldots \nu_n)$  indicate $n$-tuples of indices.
Using an analogous notation, we define the $\cdot$ product between  the generic metric  ${\bf G}_{(n)}$  and the $n$-form $\zeta$  as
 \bea
{\bf G}_{(n)}\cdot \zeta =\hat{\zeta},\eea
with 
\begin{widetext}
\bea
\hat{\zeta}_{\vec{\mu}}:=\frac{1}{n!}{[{\bf G}_{(n)}]}_{\vec{\mu}\vec{\nu}}{\zeta}_{\vec{\rho}}(dx^{\mu_1}\wedge dx^{\mu_2}\wedge\ldots dx^{\mu_n})\Avg{dx^{\nu_1}\wedge dx^{\nu_2},\wedge\ldots\wedge dx^{\nu_n},dx^{\rho_1}\wedge dx^{\rho_2}\wedge\ldots dx^{\rho_n}}.
\eea
\end{widetext}
Thus the eigenvalue problem for ${\bf G}_{(n)}$ can be written as 
\bea
{[G_{(n)}]}_{\vec{\mu}\vec{\nu}}{[g_{(n)}]}^{\vec{\nu}\vec{\rho}}\zeta_{\vec{\rho}}=\lambda\zeta_{\vec{\mu}},
\eea 
where here and in the following we have indicated with $\vec{\mu}$  $\vec{\rho}$ and $\vec{\eta}$ tuples of  $n$ indices. 
From this definition of the eigenvector of a metric between $n$-forms it follows immediately that the metric tensor ${g}_{(n)}$ has all eigenvalues $\lambda$ equal to one, i.e.
\bea
\lambda=1,
\eea
 for any arbitrary value of $n$, with $0\leq n\leq d$.
\subsection{Flattening of the  metric tensors ${\bf G}_{(n)}$ and reduction to a matrix eigenvalue problem}
To define the eigenvalues associated to the metric tensor ${\bf G}_{(1)}$ between $1$-forms, we can use a matrix formalism and define 
\bea
{[{{\bf N}}_{(1)}]}_{\mu}^{\ \nu}:={[G_{(1)}]}_{\mu\rho}g^{\rho\nu}.
\eea
The eigenvalues of the tensor ${\bf G}_{(1)}$ are defined as the eigenvalues of the matrix ${{\bf N}}_{(1)}$. Note that now the multiplication of this latter matrix with itself preserves Lorentz invariance.

The eigenvalue problem for the metric ${\bf G}_{(2)}$ Eq.(\ref{eig2_b2}) can be as well written in matrix form, by considering the $m\times m$ flattened matrices ${\bf G}_{(2)}^F$ and $g_{(2)}^F$ with $m=d(d-1)/2$ representing the dimension of the  basis for the 2-forms defined on the $d$ dimensional manifold $\mathcal{K}$.
For instance, taking the basis $\{dx^{\mu}\wedge dx^{\nu}\}$ with $\mu<\nu$ the flattened matrices ${\bf G}_{(2)}^F$ and $g_{(2)}^M$ will have respectively matrix elements 
\bea
{[G_{(2)}^F]}_{\mu\nu;\rho\sigma}=2{[G_{(2)}]}_{\mu\nu\rho\sigma},\quad {[g_{(2)}^F]}_{\mu\nu;\rho\sigma}=2{[g_{(2)}]}_{{\mu\nu\rho\sigma}}.\nonumber
\eea
A practical example might be helpful to illustrate this construction.
Assuming that the manifold $\mathcal{K}$ is $d=3$ dimensional and has three coordinates $0,1,2$, the flattened matrix ${g}_{(2)}^{F}$ associated to $g_{(2)}$, defined in the basis $dx^{0}\wedge dx^{(1)},dx^{(0)}\wedge dx^{(2)}$ and $dx^{(1)}\wedge dx^{(2)}$ is given by   
\bea
{g}_{(2)}^{F}=\begin{pmatrix}
g_{00}g_{11}-g_{01}g_{10}&g_{00}g_{12}-g_{02}g_{10}&g_{01}g_{12}-g_{02}g_{11}\\
g_{00}g_{21}-g_{01}g_{20}&g_{00}g_{22}-g_{02}g_{20}&g_{01}g_{22}-g_{02}g_{21}\\
g_{10}g_{21}-g_{11}g_{20}&g_{10}g_{22}-g_{12}g_{20}&g_{11}g_{22}-g_{12}g_{21}
\end{pmatrix}.\nonumber
\eea
Using these flattened matrices we then construct the matrix ${{\bf N}}_{(2)}$ of matrix elements
\bea
{[{{\bf N}}_{(2)}]}_{\mu\nu}^{\ \ \  \eta\theta}:={[G_{(2)}^F]}_{\mu\nu\rho\sigma}{[g_{(2)}^F]}^{\rho\sigma\eta\theta},
\eea
where $\mu<\nu$ and $\eta<\theta$.
Note that now  the usual matrix multiplication of ${{\bf N}}_{(2)}$ with itself preserves Lorentz invariance.
The eigenvalues of this matrix ${{\bf N}}_{{(2)}}$ are equal to the eigenvalue of ${\bf G}_{(2)}$ defined in Eq.(\ref{eig2_b2}).

Similarly metric tensors ${\bf G}_{(n)}$ considered as quantum operators between $n$-forms can be flattened by considering a basis of independent $n$-forms. For instance a basis for these $n$-forms can be given by the canonical $n$-forms
\bea
dx^{\mu_1}\wedge dx^{\mu_2}\wedge\ldots\wedge dx^{\mu_n},
\eea
with $\mu_1<\mu_2<\ldots<\mu_n$.
This canonical basis is formed by $m=\begin{pmatrix}
d\\n
\end{pmatrix}$ $n$-forms.

On this basis we can define the $m\times m$ flattened matrices
\bea
{[G_{(2)}^F]}_{\vec{\mu};\vec{\nu}}=n!{[G_{(2)}]}_{\vec{\mu};\vec{\nu}},\quad {[g_{(2)}^F]}_{\vec{\mu};\vec{\nu}}=n!{[g_{(2)}]}_{\vec{\mu};\vec{\nu}}.\nonumber
\eea
Starting from these flattened matrices we define the matrix ${\bf N}_{(n)}$ of elements
\bea
{[{{\bf N}}_{(n)}]}_{\vec{\mu}}^{\ \ \  \vec{\eta}}:={[G_{(n)}^F]}_{\vec{\mu}\vec{\rho}}{[g_{(n)}^F]}^{\vec{\rho}\vec{\eta}}.
\eea
As we have seen for $n=1$ and $n=2$ now these matrices can be multiplied preserving Lorentz invariance, and the eigenvalues of these matrices correspond to the eigenvalues of the metric tensor ${\bf G}_{(n)}$.

\subsection{Trace of the metric tensors ${\bf G}_{(n)}$ }
The trace of the metric tensors ${\bf G}_{(n)}$  is defined as the trace of the matrix ${{\bf N}}_{(n)}$ or equivalently as 
\bea
\mbox{Tr}{\bf G}_{(n)}=\mbox{Tr}_M{{\bf N}}_{(n)}.
\eea
For ease of notation in the following we will also define  indicate this trace with
\bea
\mbox{Tr}_F {\bf G}_{(n)}g_{(n)}^{-1}=\mbox{Tr}{\bf G}_{(n)}=\mbox{Tr}_M{{\bf N}}_{(n)}.
\eea
From the above definition of ${\bf N}_{(n)}$ it follows that our definition of the trace is the usual definition of the trace of a tensor, i.e., 
\bea
\mbox{Tr}_F{\bf G}_{(n)}g_{(n)}^{-1}=
{[G_{(n)}]}_{\vec{\mu}\vec{\rho}}{[g_{(n)}]}^{\vec{\rho}\vec{\mu}}.
\eea
Specifically,  for  $n\in \{1,2\}$ we obtain
\bea
\mbox{Tr}{\bf G}_{(1)}&=&{[G_{(1)}]}_{\mu\rho}g^{\rho\mu},\nonumber \\
\mbox{Tr}{\bf G}_{(2)}&=&{[G_{(2)}]}_{\mu\nu\rho\sigma}{[g_{(2)}]}^{\rho\sigma\mu\nu},
\eea
and for general $n$ we have 
It follows that the trace of   $g_{(n)}$ are given by 
\bea
\mbox{Tr}{\bf g}_{(n)}=\begin{pmatrix}
d\\n
\end{pmatrix},
\eea
for every $0\leq n\leq d$.

\subsection{Entropy for metric tensors  ${\bf G}_{(n)}$}

The entropy $H$ of metric tensor ${\bf G}_{(n)}$ is defined as 
\bea
H=\mbox{Tr}{\bf G}_{(n)}\ln {\bf G}_{(n)}^{-1}:=-\sum_{\lambda}\lambda\ln \lambda
\eea
where $\lambda$ indicates the generic eigenvalue of ${{\bf G}_{(n)}}$.
Since $g_{(n)}$ has all eigenvalues equal to one, it follows 
\bea
H_{(n)}=\mbox{Tr}{ g}_{(n)}\ln {g}_{(n)}^{-1}=0, 
\label{gntrace}
\eea
for any value of $0\leq n\leq d$.
The cross entropy between $g_{(n)}$ and ${\bf G}_{(n)}$
is defined as 
\bea
\mbox{Tr}g_{(n)}\ln {\bf G}_{(n)}^{-1} &:=&-\mbox{Tr}_F \ln {\bf G}_{(n)} g^{-1}_{(n)}
\nonumber \\&:=&-\mbox{Tr}_M{{\bf N}}_{(n)}=-\sum_{\mu}\ln (\mu),
\eea
where $\mu$ indicates the generic eigenvalue of ${\bf N}_{(n)}$.
%Specifically, we have that for $n=1$
%\bea
%\mbox{Tr}{g_{(1)}\ln {\bf G}_{(1)}^{-1}&:=&-\mbox{Tr}_M\ln {\bf N}_{(1)}\nonumber \\&=&-\mbox{Tr}_M \ln {\bf G}_{(1)}g^{-1}.
%\eea

\section{The topological  metric as a quantum operator}
\label{quantum}
{  
%Before investigating the properties of  the  equations of motion associated to our entropic action, 
In this Appendix  we outline the underlying quantum  theory of the proposed general scenario and the relation of the proposed entropy action with the Araki quantum relative entropy~\cite{araki1975relative,ohya2004quantum}.
%At first glance, readers primarily interested in modified gravity may choose to skip this paragraph.

Our treatment will consider quantum operators in first quantization. We  will provide the foundation of  the theory of topological metric tensors interpreted as quantum operators \cite{araki1999mathematical}  where the vectors of our Hilbert space are the topological fields formed by the direct sum of  a $0$-form, a $1$-form and a $2$-form.

We consider the two generic topological fields $\ket{\Psi}$ and $\ket{\Phi}$ given by 
\bea
\ket{\Psi}={{\phi}}\oplus {\omega}_{\mu} dx^{\mu}\oplus{\zeta}_{\mu\nu} dx^{\mu}\wedge dx^{\nu},\nonumber \\
\ket{\Phi}=\hat{\phi}\oplus \hat{\omega}_{\mu} dx^{\mu}\oplus \hat{\zeta}_{\mu\nu} dx^{\mu}\wedge dx^{\nu}.\label{phi1}
\eea
The scalar product among these two topological fields is given by 
\bea
\Avg{\langle \Psi,\Phi\rangle}=\int \sqrt{-|g|} \Big(\bar{\phi}\hat{\phi}+\bar{\omega}_{\mu}\hat{\omega}^{\mu}+\bar{\zeta}_{\mu\nu}\hat{\zeta}^{\mu\nu}\Big) d{\bf r},
\eea
where $\hat\omega^{\mu}=g^{\mu\rho}\hat\omega_{\rho}$ and $\hat\zeta^{\mu\nu}=g^{\mu\rho}g^{\nu\sigma}\hat\zeta_{\rho\sigma}={[g_{(2)}]}^{\mu\nu\rho\sigma}\hat\zeta_{\rho\sigma}$.
Thus the Hilbert space $\mathcal{H}$ is endowed with a scalar product defined through the default metric tensor $\tilde{g}^{-1}$ given by 
\bea
\tilde{g}^{-1}&=&1\oplus g^{\mu\nu}dx_{\mu}\otimes dx_{\nu}\nonumber \\
&&\oplus {[g_{(2)}]}^{\mu\nu\rho\sigma}\Big(dx_{\mu}\wedge dx_{\nu}\Big)\otimes \Big(dx_{\rho}\wedge dx_{\sigma}\Big)
\eea
This scalar product has the following properties:
\begin{itemize}
\item {\it Linearity on the second term}
\bea
\Avg{\langle{\Psi},c_1\Phi_1+c_2\Phi_2\rangle}=c_1
\Avg{\langle{\Psi},\Phi_1\rangle}+c_2 \Avg{\langle{\Psi},\Phi_2\rangle},
\eea
for arbitrary $c_1,c_2\in \mathbb{C}$.
\item {\it Antilinearity on the first term}
\bea
\Avg{\langle{c_1\Psi_1+c_2\Psi_2}\Phi\rangle}=\bar{c}_1\Avg{\bra{\Psi_1}\Phi\rangle}+\bar{c}_2\Avg{\bra{\Psi_2}\Phi\rangle},
\eea
for arbitrary $c_1,c_2\in \mathbb{C}$.

\item{\it Scalar product of a topological field with itself}
The scalar product of a topological field with itself is real:
\bea
\Avg{\langle{\Psi},\Phi\rangle}
\in \mathbb{R},
\eea
unless is infinite.
\end{itemize}
The Hilbert space $\mathcal{H}$ is formed by all topological fields $\ket{\Phi}$ such that
\bea
\Avg{\bra{\Phi}\Phi\rangle}<\infty.
\label{hilb}
\eea 

The metric induced by the matter field $\tilde{\bf G}$ 
\bea
\tilde{\bf G}&=&{[G_{(0)}]}\oplus {[G_{(1)}]}_{\mu\nu}dx^{\mu}\otimes dx^{\nu}\nonumber \\
&&\oplus{[G_{(2)}]}_{\mu\nu\rho\sigma}\Big(dx^{\mu}\wedge dx^{\nu}\Big)\otimes \Big(dx^{\rho}\wedge dx^{\sigma}\Big).
\eea
endowed with the  $\cdot$ product defined in the Appendix \ref{ApA} (Eq.(\ref{cdot1}) and (\ref{cdot2})) can be viewed as a quantum operator $\tilde{\bf G}:\mathcal{H}\to \mathcal{H}$ where  $\tilde{\bf G}\cdot \ket{\Phi}\in \mathcal{H}$,  
\bea
\tilde{\bf G}\cdot \ket{\Phi}&=&\phi\oplus {[G_{(1)}]}_{\mu\nu}\omega^{\nu}dx^{\mu}\nonumber \\
&&\oplus{[G_{(2)}]}_{\mu\nu\rho\sigma}\zeta^{\rho\sigma}dx^{\mu}\wedge dx^{\nu}.
\eea

The metric $\tilde{g}$ of the manifold $\mathcal{K}$ can be used to define a dual Hilbert space $\mathcal{H}^{\star}$ given by the dual topological field. For instance  $\ket{\Psi^{\star}}$ is the dual of $\ket{\Psi}$ and $\ket{\Phi^{\star}}$ is the dual of $\ket{\Phi}$ with $\ket{\Psi^{\star}},\ket{\Phi^{\star}}$ given by
\bea
\ket{\Psi^{\star}}={{\phi}}\oplus {\omega}^{\mu} dx_{\mu}\oplus{\zeta}^{\mu\nu} dx_{\mu}\wedge dx_{\nu},\nonumber \\
\ket{\Phi^{\star}}=\hat{\phi}\oplus \hat{\omega}^{\mu} dx_{\mu}\oplus \hat{\zeta}^{\mu\nu} dx_{\mu}\wedge dx_{\nu},
\eea
where $\omega^{\mu}=g^{\mu\rho}\omega_{\rho},\hat{\omega}^{\mu}=g^{\mu\rho}\hat{\omega}_{\rho}$ and $\zeta^{\mu\nu}={[g_{(2)}]}^{\mu\nu\rho\sigma}\zeta_{\rho\sigma},\hat{\zeta}^{\mu\nu}={[g_{(2)}]}^{\mu\nu\rho\sigma}\hat\zeta_{\rho\sigma}$.
While the Hilbert space $\mathcal{H}$ is endowed with a scalar product defined through the default metric tensor $\tilde{g}^{-1}$
the Hilbert space $\mathcal{H}^{\star}$ is endowed with a scalar product $\Avg{\Avg{ \Psi^{\star},\Phi^{\star}}}_{\star}$ mediated by $\tilde{g}$ such that
\bea
\Avg{\langle{\Psi},\Phi\rangle}=\Avg{\langle \Psi^{\star},\Phi^{\star}\rangle}_{\star}.
\label{Hdual}
\eea 

The dual operator $\tilde{\bf G}^{\star}:\mathcal{H}^{\star}\to\mathcal{H}^{\star}$ of $\tilde{\bf G}$  is given by 
\bea
\tilde{\bf G}^{\star}&=&{G_{(0)}^{\star}}\oplus {[G_{(1)}^{\star}]}^{\mu\nu}dx_{\mu}\otimes dx_{\nu}\nonumber \\
&&\oplus{[G_{(2)}^{\star}]}^{\mu\nu\rho\sigma}\Big(dx_{\mu}\wedge dx_{\nu}\Big)\otimes \Big(dx_{\rho}\wedge dx_{\sigma}\Big).
\eea
and satisfies
\bea
\Avg{\Avg{{\Psi},\tilde{\bf G}\cdot\Phi}}=\Avg{\Avg{ \tilde{\bf G}^{\star}\cdot\Psi^{\star},\Phi^{\star}}}_{\star},
\label{HdualGs}
\eea 
for any arbitrary choice of $\ket{\Psi}$ and $\ket{\Phi}$. Here the $\cdot$ product of $\tilde{\bf G}^{\star}$ with $\ket{\Phi^{\star}}$ is given by 
\bea
\tilde{\bf G}^{\star}\cdot \ket{\Phi^{\star}}&=&{[G_{(0)}^{\star}]}\phi\oplus {[G_{(1)}^{\star}]}^{\mu\nu}\omega_{\nu}dx_{\mu}\nonumber \\
&&\oplus{[G_{(2)}^{\star}]}^{\mu\nu\rho\sigma}\zeta_{\rho\sigma}dx_{\mu}\wedge dx_{\nu}.
\eea
Due to the fact the the metrics ${\bf G}_{(n)}$ that constitute $\tilde{\bf G}$ are Hermitian under the exchange of the first $n$ and the second $n$ indices, we have that $\tilde{\bf G}^{\star}$ is related to $\tilde{\bf G}$ by
\bea
G_{(0)}^{\star}&=&G_{(0)},\nonumber \\ {[G_{(1)}^{\star}]}^{\mu\nu}&=&{[G_{(1)}]}^{\mu\nu},\nonumber \\ 
{[G_{(2)}^{\star}]}^{\mu\nu\rho\sigma}&=&{[G_{(1)}]}^{\mu\nu\rho\sigma},
\eea
where 
\bea
{[G_{(1)}]}^{\mu\nu}&=&g^{\mu\rho}{[G_{(1)}]}_{\rho\sigma}g^{\nu\sigma},\nonumber \\ 
{[G_{(2)}]}^{\mu\nu\rho\sigma}
&=&{[g_{(2)}]}^{\mu\nu\eta_1\eta_2}{[G_{(2)}]}_{\eta_1\eta_2\theta_1\theta_2}{[g_{(2)}]}^{\theta_1\theta_2\rho\sigma}.\nonumber 
\eea
Thus we will indicate the relation between $\tilde{\bf G}^{\star}$ and $\tilde{\bf G}$ for short as 
\bea
\tilde{\bf G}^{\star}=\tilde{g}^{-1}\tilde{\bf G}\tilde{g}^{-1}.
\eea
Form this relation it follows that  the dual of the default metric $\tilde{g}$ is the inverse metric $\tilde{g}^{-1}$, i.e.
\bea
\tilde{g}^{\star}=\tilde{g}^{-1}.
\eea
Interestingly, the metric tensors are such that the dual of the dual coincides with the metric tensor itself, indeed we have
\bea
\tilde{\bf G}=\tilde{\bf G}^{\star\star}=\tilde{g}\tilde{\bf G}^{\star}\tilde{g}.
\eea

The topological metrics  $\tilde{\bf G}$ and $\tilde{\bf G}^{\star}$ that we consider in this work are respectively elements of  the algebras $\textswab{U}$ and $\textswab{U}^*$ that generalizes the $C^*$ algebra \cite{araki1999mathematical} and have the following properties:
\begin{itemize}
\item[(a)] $\textswab{U}$ and $\textswab{U}^{\star}$ are  algebras with complex numbers as the coefficient field.
\item[(b)] The product between $\tilde{\bf G}_1, \tilde{\bf G}_2\in \textswab{U}$  given by 
\bea
\tilde{\bf G}_{n}&=&G_{(0),n}\oplus[G_{(1),n}]_{\mu\nu}dx^{\mu}\otimes dx^{\nu}\nonumber \\
&&\oplus [G_{[2],n}]_{\mu\nu\rho\sigma}(dx^{\mu}\wedge dx^{\nu})\otimes(dx^{\rho}\wedge dx^{\sigma}),
\eea
for $n\in \{1,2\},$
is mediated by the metric $\tilde{g}^{-1},$ while the product between $\tilde{\bf G}_1^{\star}, \tilde{\bf G}_2^{\star}\in \textswab{U}^{\star}$ is mediated by the metric $\tilde{g}.$ Specifically we have,
\bea
&&\tilde{\bf G}_1 \tilde{g}^{-1} \tilde{\bf G}_2={[G_{(0),12}]}\oplus {[G_{(1),12}]}_{\mu\nu}dx^{\mu}\otimes dx^{\nu}\nonumber \\
&&\hspace{4mm}\oplus{[G_{(2),12}]}_{\mu\nu\rho\sigma}\Big(dx^{\mu}\wedge dx^{\nu}\Big)\otimes\Big(dx^{\rho}\wedge dx^{\sigma}\Big),\nonumber
\eea
with
\bea
{[G_{(0),12}]}&=&{[G_{(0),1}]}{[G_{(0),2}]},\nonumber \\
{[G_{(1),12}]}_{\mu\nu}&=&{[G_{(1),1}]}_{\mu\eta_1}g^{\eta_1\eta_2}{[G_{(1),2}]}_{\eta_2\nu},\nonumber\\
{[G_{(2),12}]}_{\vec{\mu}\vec{\rho}}&=&{[G_{(2),1}]}_{\vec{\mu}\vec{\eta}_1}{[g_{(2)}]}^{\vec{\eta}_1,\vec{\eta}_2}{[G_{(2),2}]}_{\vec{\eta}_2\vec{\rho}}.
\eea
Similarly we have
\bea
\tilde{\bf G}_1^{\star} \tilde{g} \tilde{\bf G}_2^{\star}&=&{[G_{(0),12}^{\star}]}\oplus {[G_{(1),12}^{\star}]}^{\mu\nu}dx_{\mu}\otimes dx_{\nu}\nonumber \\
&&\oplus{[G_{(2),12}^{\star}]}^{\mu\nu\rho\sigma}\Big(dx_{\mu}\wedge dx_{\nu}\Big)\otimes\Big(dx_{\rho}\wedge dx_{\sigma}\Big),\nonumber
\eea
with
\bea
{[G_{(0),12}^{\star}]}&=&{[G_{(0),1}^{\star}]}{[G_{(0),2}^{\star}]},\nonumber \\
{[G_{(1),12}^{\star}]}^{\mu\nu}&=&{[G_{(1),1}^{\star}]}^{\mu\eta_1}g_{\eta_1\eta_2}{[G_{(1),2}^{\star}]}^{\eta_2\nu},\nonumber\\
{[G_{(2),12}^{\star}]}^{\vec{\mu}\vec{\rho}}&=&{[G_{(2),1}^{\star}]}_{\vec{\mu}\vec{\eta}_1}{[g_{(2)}]}_{\vec{\eta}_1,\vec{\eta}_2}{[G_{(2),2}^{\star}]}^{\vec{\eta}_2\vec{\rho}}.
\eea
\item[(c)] A bijection $\tilde{\bf G}\in \textswab{U} \to \tilde{\bf G}^{\star}\in \textswab{U}^{\star}$ and a bijection $\tilde{\bf G}^{\star}\in \textswab{U}^{\star} \to \tilde{\bf G}^{\star\star}\in \textswab{U}$ are defined and satisfy the following properties:
\begin{itemize}
\item[(i)]
The dual of the dual  coincides with the metric tensor itself,
\bea
\Big(\tilde{\bf G}^{\star}\Big)^{\star}=\tilde{\bf G}^{\star\star}=\tilde{\bf G}.
\eea
\item[(ii)]
The product between metrics in $\textfrak{U}$ maps, under the bijection, to the product between dual metrics in $\textfrak{U}^{\star}$ and vice-versa according to:
\bea
\Big(\tilde{\bf G}_1 \tilde{g}^{-1} \tilde{\bf G}_2\Big)^{\star}&=&\tilde{\bf G}_2^{\star} \tilde{g} \tilde{\bf G}_1^{\star},\nonumber \\ \Big(\tilde{\bf G}_1^{\star} \tilde{g} \tilde{\bf G}_2^{\star}\Big)^{\star}&=&\tilde{\bf G}_2 \tilde{g}^{-1} \tilde{\bf G}_1.\eea
\item[(iii)]
The dual of a linear combination of metrics is $\textfrak{U}$ and the dual of a linear combination of metrics in $\textfrak{U}^{\star}$ obeys:
\bea
(c_1\tilde{\bf G}_1+c_2\tilde{\bf G}_2)^{\star}&=&\bar{c}_1\tilde{\bf G}_1^{\star}+\bar{c}_2\tilde{\bf G}_2^{\star},\nonumber \\
(c_1\tilde{\bf G}_1^\star+c_2\tilde{\bf G}_2^{\star})^{\star}&=&\bar{c}_1\tilde{\bf G}_1+\bar{c}_2\tilde{\bf G}_2. 
\eea
\end{itemize}
\item[(d)] The norm associated to the topological metric $\tilde{\bf G}\in \textfrak{U}$ is   equal to the norm associated to the dual 
and given by $\|\tilde{\bf G}\|=\|\tilde{\bf G}^{\star}\|$ defined as
\bea
\|\tilde{\bf G}\|=\|\tilde{\bf G}^{\star}\|=\int \sqrt{|-g|}\mbox{Tr}_F \Big(\tilde{\bf G}\tilde{\bf G}^{\star}\Big) d{\bf r},
\label{norm}
\eea
where $\mbox{Tr}_F \Big(\tilde{\bf G}\tilde{\bf G}^{\star}\Big)$ is given by  
\bea
\mbox{Tr}_F \Big(\tilde{\bf G}\tilde{\bf G}^{\star}\Big)&=&G_{(0)}^2+{[G_{(1)}]}_{\mu\nu}{[G_{(1)}]}^{\nu\mu}\nonumber \\&&+{[G_{(2)}]}_{\mu\nu\rho\sigma}{[G_{(2)}]}^{\rho\sigma\mu\nu}.
\eea
%Thus the topological metrics are hereby interpreted as bounded quantum operators. 
\end{itemize}

For a topological metric $\tilde{\bf G}$  interpreted as a quantum operator in $\textfrak{U}$ we define the square root of the modular operator $\bm\Delta^{1/2}_{\tilde{\bf G},g}:\mathcal{H}\to \mathcal{H}$ as 
\bea
\bm\Delta^{1/2}_{\tilde{\bf G},g}=\sqrt{\tilde{\bf G}\tilde{\bf G}^{\star}}=\tilde{\bf G}\tilde{g}^{-1},
\eea
where the last identity is derived under the assumption that $\tilde{\bf G}$ is positively definite, i.e. it has only positive eigenvalues.
In this assumption the action of the  square root of the modular operator $\bm\Delta_{\tilde{\bf G},g}^{1/2}$ on the topological field $\ket{\Phi}$ given by Eq.(\ref{phi1}) can be explicitly expressed as  
\bea\bm \Delta^{1/2}_{\tilde{\bf G},g}\ket{\Phi}&=&{G_{(0)}}\phi\oplus {[G_{(1)}]}_{\mu\rho}g^{\rho\nu}\omega_{\nu}dx^{\mu}\nonumber \\
&&\hspace{-5mm}\oplus{[G_{(2)}]}_{\mu\nu\eta_1\eta_2}g_{(2)}^{\eta_1\eta_2\rho\sigma}\zeta_{\rho\sigma}\Big(dx^{\mu}\wedge dx^{\nu}\Big).
\eea
In terms of the modular operator the considered entropic action is defined similarly to the usual Araki quantum relative entropy \cite{araki1975relative,ohya2004quantum} as 
\bea
\mathcal{S}=\frac{1}{\ell_P}\int \sqrt{|-g|}\mathcal{L} d{\bf r},
\eea
where \bea
\mathcal{L}=-\mbox{Tr}_F\ln \bm\Delta^{1/2}_{\tilde{\bf G},g}=-\mbox{Tr}_F\ln \tilde{\bf G}\tilde{g}^{-1}.
\eea
Thus the  entropic action adopted in this work is strictly related to the Araki quantum relative entropy~\cite{araki1975relative,ohya2004quantum}, albeit the definition needs to take into account the structure of the Hilbert space $\mathcal{H}$ that is here formed by topological fields given by the direct sum of  a $0$-form, with a $1$-form and a $2$-form.
}
\bibliographystyle{unsrt}
\bibliography{references}

\end{document}